\documentclass[12pt]{article}
\usepackage[english]{babel}
\usepackage[a4paper,bindingoffset=0.0in,
left=0.75in,right=0.75in,top=1in,bottom=1in,footskip=.25in]{geometry}

\usepackage{array}
\usepackage[utf8]{inputenc}

\usepackage{graphicx}
\usepackage{caption}
\usepackage{subcaption}

\usepackage{multirow}
\usepackage{tabularx}

\usepackage{titling}
\newcommand{\subtitle}[1]{%
  \posttitle{%
    \par\end{center}
    \begin{center}\large#1\end{center} 
    \vskip0.1em}%
}
\usepackage{amsmath}

\usepackage{authblk}

\usepackage[bottom]{footmisc}
\usepackage[colorlinks,citecolor=blue,urlcolor=blue,linkcolor=blue]{hyperref}

\begin{document}
\title{\textbf{Describing the Structural Phenotype of the Glaucomatous Optic Nerve Head Using Artificial Intelligence}}
\date{}

\author[1,2]{Satish K. Panda}
\author[1,2]{Haris Cheong}
\author[3]{Tin A. Tun}
\author[1,2]{Sripad K. Devella}
\author[4]{Ramaswami Krishnadas}
\author[2]{Martin L. Buist}
\author[3,5]{Shamira Perera}
\author[3,7]{Ching-Yu Cheng}
\author[3,7]{Tin Aung}
\author[6 ]{Alexandre H. Thi{\'e}ry}
\author[1,5]{Micha{\"e}l J. A. Girard}

\affil[1]{Ophthalmic Engineering \& Innovation Laboratory (OEIL), Singapore Eye Research Institute, Singapore National Eye Centre, Singapore}
\affil[2]{Department of Biomedical Engineering, National University of Singapore, Singapore}
\affil[3]{Singapore Eye Research Institute, Singapore National Eye Centre, Singapore}
\affil[4]{Glaucoma Services, Aravind Eye Care Systems, Madurai, India}
\affil[5]{Duke-NUS Medical School, Singapore}
\affil[6]{Department of Statistics and Applied Probability, National University of Singapore, Singapore}
\affil[7]{Department of Ophthalmology, Yong Loo Lin School of Medicine, National University of Singapore, Singapore}

\date{}                     
\setcounter{Maxaffil}{0}
\renewcommand\Affilfont{\itshape\small}

\maketitle

\footnotetext[1]{Correspondence: Micha{\"e}l J. A. Girard, Ophthalmic Engineering and Innovation Laboratory,  Singapore Eye Research Institute, 20 College Road Discovery Tower, The Academia, 169856, Singapore. \\email: {mgirard@ophthalmic.engineering}}

\begin{abstract}
The optic nerve head (ONH) typically experiences complex neural- and connective-tissue structural changes with the development and progression of glaucoma, and monitoring these changes could be critical for improved diagnosis and prognosis in the glaucoma clinic. The gold-standard technique to assess structural changes of the ONH clinically is optical coherence tomography (OCT). However, OCT is limited to the measurement of a few hand-engineered parameters, such as the thickness of the retinal nerve fiber layer (RNFL), and has not yet been qualified as a stand-alone device for glaucoma diagnosis and prognosis applications. We argue this is because the vast amount of information available in a 3D OCT scan of the ONH has not been fully exploited. In this study we propose a deep learning approach that can: \textbf{(1)} fully exploit information from an OCT scan of the ONH; \textbf{(2)} describe the structural phenotype of the glaucomatous ONH; and that can \textbf{(3)} be used as a robust glaucoma diagnosis tool. Specifically, the structural features identified by our algorithm were found to be related to clinical observations of glaucoma. The diagnostic accuracy from these structural features was $92.0 \pm 2.3 \%$ with a sensitivity of $90.0 \pm 2.4 \% $ (at $95 \%$ specificity). By changing their magnitudes in steps, we were able to reveal how the morphology of the ONH changes as one transitions from a `non-glaucoma' to a `glaucoma' condition. We believe our work may have strong clinical implication for our understanding of glaucoma pathogenesis, and could be improved in the future to also predict future loss of vision.
\end{abstract}

\section{Introduction}
\paragraph{}
The central event in glaucoma is slow and irreversible damage of retinal ganglion cell axons \cite{osborne1999ganglion,weinreb2014pathophysiology}. Retinal ganglion cells carry visual information from the retina to the brain. When damaged, they undergo programmed cell death (apoptosis) resulting in vision loss. The main site of retinal ganglion cell damage occurs within the optic nerve head (ONH) located at the back of the eye \cite{quigley1981optic}. To date, it is widely accepted that clinical evaluation and documentation of the ONH is essential for the diagnosis and the monitoring of glaucoma \cite{conlon2017glaucoma, tatham2014strategies}.

However, providing an accurate glaucoma diagnosis is a complex endeavor that is time consuming for clinicians. It is subjective and heavily dependent on a clinician's experience/expertise, and in more complex cases, requires a battery of multiple clinical tests. Such tests often need to be repeated at multiple patients' visits to overcome their inherent subjectivity and patient variability to confirm a diagnosis. As stated by the World Glaucoma Association: ``As yet there is no widely-accepted method of combining the results of several [glaucoma] tests [to provide an improved diagnosis]''. In fact, the use of multiple glaucoma diagnostic tests has been found to increase the likelihood of false-positives yielding to over-treatment \cite{weinreb2017diagnosis}.

Axonal damage in glaucoma is typically evaluated through two tests that assess ONH structure and visual function. In routine clinical practice, functional measures of axonal loss are frequently assessed with visual field testing. During testing, flashes of light reach various regions of the retina to generate a map of its sensitivity to light. Structural measures of axonal loss are assessed with optical coherence tomography (OCT) - a 3D imaging modality that allows fast, high-resolution and non-invasive visualization of the ONH \cite{li2016imaging}. Using OCT, studies have found that retinal nerve fiber layer (RNFL) thinning was strongly associated with glaucoma severity \cite{bowd2000retinal, christopher2018retinal, kim2009retinal, oddone2016macular}. Because RNFL thinning generally precedes detectable functional defects in glaucoma patients \cite{banitt2013progressive}, RNFL thickness (assessed through OCT) has remained a gold-standard parameter for glaucoma.

However, Li and Jampel concluded that: ``OCT was not a sufficient stand-alone test for the detection of glaucoma or for triage use in primary care'' \cite{li2016imaging}. We argue this is because the vast amount of information available in a 3D OCT scan of the ONH has not been fully exploited. RNFL thickness and the more recent minimum-rim-width are 1-dimensional (depth) parameters representative of neural tissues only. However, connective tissues also exhibit complex 3-dimensional changes during the development and progression of glaucoma, including, but not limited to: post-laminar deformations \cite{ethier2006predicted}, changes in LC and choroidal thickness \cite{usui2012evaluation, vianna2017serial}, peripapillary atrophy \cite{jonas2005clinical}, scleral canal expansion \cite{bellezza2003deformation}, posterior migration of the LC \cite{lee2018comparison}, and peripapillary scleral bowing \cite{wang2020peripapillary}. To date, no clinical solutions currently consider changes in both neural and connective tissues simultaneously, which may also explain the lack in diagnostic power.

Recently, several artificial intelligence (AI) studies have proposed deep learning algorithms applied to OCT images of the ONH to provide an improved glaucoma diagnosis \cite{lavinsky2017future,maetschke2019feature, medeiros2019deep, ran2020deep}. Most have reported encouraging performance, with or without the inclusion of additional demographic/clinical parameters and tests (e.g., visual field maps). However, none has attempted to provide a fundamental understanding of what should be the structural signature of a glaucomatous ONH. In other words, why is an AI algorithm capable of classifying `glaucoma' from `non-glaucoma' from an OCT scan of the ONH? While it may sound like a trivial question, further research is actually needed.

In this study, we aim to use AI to describe the structural phenotype of the glaucomatous ONH using a deep learning approach known as autoencoder. Autoencoders have gained popularity in recent years for `feature learning' applications, because they can learn low dimensional features from high-dimensional datasets \cite{berchuck2019scalable, berchuck2019estimating}.

\section{Methods}

\subsection{Patient recruitment and OCT imaging}
\paragraph{}
A total of 3,782 subjects (2,233 glaucoma and 1,549 non-glaucoma) were recruited for this study at two different centers: the Singapore National Eye Center (SNEC, Singapore) and the Aravind Eye Hospital (Madurai, India). The study at SNEC had three different cohorts, i.e., Cohort 1: 1,008 subjects (703 glaucoma and 305 non-glaucoma) of Chinese ethnicity, Cohort 2: 894 subjects (53 glaucoma and 841 non-glaucoma) of Indian ethnicity, and Cohort 3: 415 subjects (27 glaucoma and 388 non-glaucoma) of Chinese ethnicity. On the other hand, the study at Aravind Eye Hospital had 1,465 patients (1,450 glaucoma and 15 non-glaucoma) of Indian ethnicity only (see Table 1 for further details). All subjects gave written informed consent. The study adhered to the tenets of the Declaration of Helsinki and was approved by the institutional review board of the respective hospitals. Subjects with intraocular pressure (IOP) less than 21 mmHg, healthy optic discs with a vertical cup-disc ratio (VCDR) less than or equal to 0.5, and normal visual fields tests were considered as non-glaucoma, whereas subjects with glaucomatous optic neuropathy, VCDR $>0.7$, and/or neuroretinal rim narrowing with repeatable glaucomatous visual field defects were considered as glaucoma. Subjects with corneal abnormalities that have the potential to preclude the quality of the scans were excluded from the study.

\begin{center}
\begin{table}[htp]
\centering
\begin{tabular}{ |p{2.5cm}|p{1.7cm}|p{2cm}|p{1.1cm}|p{1.5cm}|p{1.7cm}|c| } 
 \hline
 Institute & Study & Age (mean $\pm$ SD) & Sex (\% of male)  & Non-glaucoma scans & Glaucoma scans & Total \\ 
 \hline
 \multirow{3}{2.5cm}{Singapore Eye Research Institute} & Cohort 1 & 64.3 $\pm$ 7.1 & 49  & 788 & 1006 & 1794 \\ 
 & Cohort 2 & 59.6 $\pm$ 9.93 & 51  & 1317 & 19 & 1336 \\ 
 & Cohort 3 & 57.7 $\pm$ 10.1 & 50  & 2062 & 12 & 2074 \\ 
 \hline
 Aravind Eye Hospital, \quad India & Cohort 1 & 56.8 $\pm$ 11.8 & 75 & 32 & 2241 & 2273 \\ 
 \hline
 & & & & 4253 & 3278 & 7531 \\
\hline
\end{tabular}
\caption{A summary of patient populations}
\label{table:Population data}
\end{table}
\end{center}

We used a standard spectral-domain OCT system (Spectralis; Heidelberg Engineering, Heidelberg, Germany) to scan left/right eye of each patient. During the scanning, patients were seated in a dark room and imaged using a single operator at each center. Each OCT volume obtained from patients consisted of 97 horizontal B-scans ($32 \mu m$ distance between B-scans, 384 A-scans per B-scan), covering a rectangular area of $15^{0} \times 10^{0}$ centered on the ONH. In total, we obtained 7,531 scans (3,278 glaucoma and 4,253 non-glaucoma scans) for our study.

\subsection{Segmenting neural and connective tissues in OCT images of the ONH}
\begin{figure}
     \centering
     \begin{subfigure}[b]{0.45\textwidth}
         \centering
         \includegraphics[height = 8cm, width = 6cm]{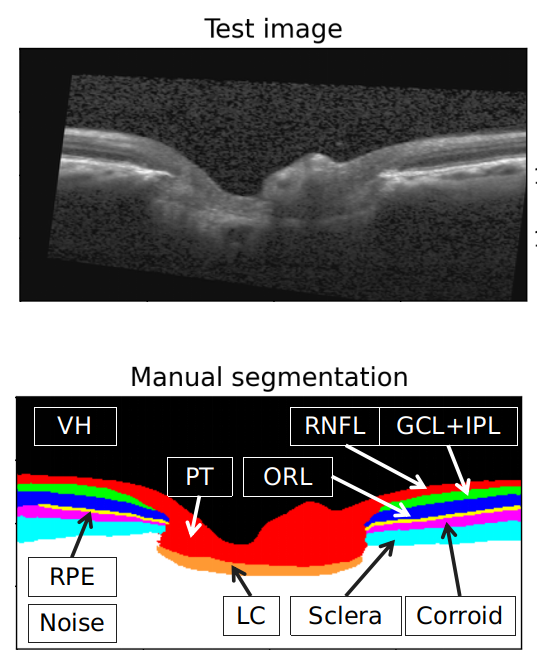}
         \caption{}
         \label{fig:AmiraSegmentation}
     \end{subfigure}
     \begin{subfigure}[b]{0.45\textwidth}
         \centering
         \includegraphics[height = 8cm, width = 9cm]{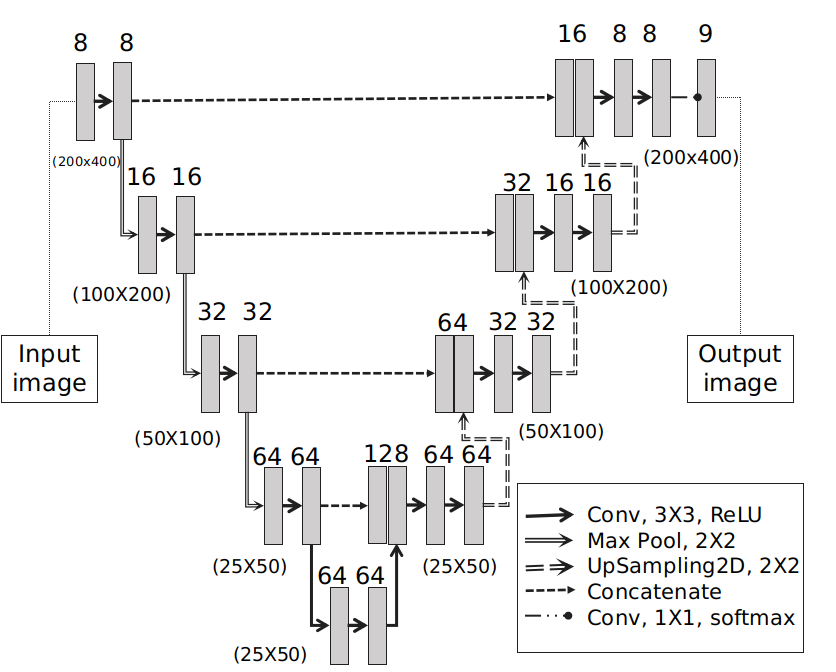}
         \caption{}
         \label{fig:UNet}
     \end{subfigure}
        \caption{a) An example OCT image (top) with the corresponding manual segmentation (bottom) b) The architecture of the U-net network.}
\end{figure}

\paragraph{}
Describing the structural signature of the glaucomatous ONH strictly using raw OCT scans is challenging because of the inherent speckle noise, blood vessel shadows, poor tissue structure visibility, and intensity inhomogeneity (Fig.~\ref{fig:AmiraSegmentation}, Top) \cite{cheong2020deshadowgan, girard2015enhancement}. An effective way to address this issue is to, in a first step, segment and thus identify or label all neural and connective tissue layers of the ONH (Fig.~\ref{fig:AmiraSegmentation}, Bottom) using a deep learning network as was performed in our previous work \cite{devalla2018drunet, DevallaTowards, devalla2018deep}. Using segmentation as a first step will facilitate our understanding of complex structural differences between glaucoma and non-glaucoma eyes.

In recent years, some studies have featured the use of CNN in a U-net architecture for image segmentation \cite{ronneberger2015u}. It exploits the advantages of skip connections and offers robust segmentations with a minimal number of trainable parameters. Although there exist other segmentation networks, e.g., DRUNET \cite{devalla2018drunet}, capsule network \cite{lalonde2018capsules}, CNN in DenseNet architecture \cite{pekala2019deep}, we used U-net for our study as it is very robust, easy-to-design and train. We optimized the U-net architecture to make it suitable for OCT image segmentation. The customized network had only 321,777 parameters as compared to 31,031,685 parameters in its original implementation \cite{ronneberger2015u}. Fig.~\ref{fig:UNet} shows the architecture of the new network, and in the appendix, a detailed description of the network is presented.

To train the segmentation network, we manually segmented 200 OCT images using Amira (version 5.4, FEI, Hillsboro, OR). We chose 50 images from each study listed in Table \ref{table:Population data}, and then split the dataset into training ($70\%$), validation ($15\%$), and test ($15\%$) sets, respectively. The split was performed in such a way that there would not be any duplicate images or images from the same subject in different sets. For some OCT images, we were unable to obtain a full-thickness segmentation of the peripapillary sclera and the LC due to limited visibility, and in such cases, only the visible portions were segmented.

To reduce computational complexity during the training process, the OCT images were resized to $200\times 400$ pixels before manual segmentation and the network was trained to delineate and extract seven layers. Specifically, we segmented the following layers: 1) the vitreous humor (VH), in black; 2) the RNFL and the prelamina (PT), in red; 3) the ganglion cell layer and inner plexiform layer (GCL+IPL), in green; 4) all other retinal layers (ORL), in blue; 5) the retinal pigment epithelium (RPE), in yellow; 6) the choroid, in pink; 7) the peripapillary sclera, in cyan; 8) the LC, in orange; 9) noise, in white. As we marked the VH and the noise below the peripapillary sclera with different labels, our segmented OCT image had in total nine different labels (Fig.~\ref{fig:AmiraSegmentation}).

To evaluate the accuracy of the segmentation network, we calculated the Dice coefficients by comparing the network predicted labels with the corresponding manually segmented images from the test set. The Dice coefficient is a metric to evaluate semantic segmentation (see appendix for mathematical descriptions). A value of one for this metric indicates a perfect match between the network prediction and the manual segmentation, and a value of zero indicates no overlap.

\subsection{Use of an autoencoder to describe the structural signature of the glaucomatous ONH}
\paragraph{}
Autoencoders are neural networks that can be used to learn low-dimensional features from a high-dimensional dataset \cite{ng2011sparse}. They usually consist of an encoder that reduces the dimension of a dataset to a small number of latent variables, and a decoder that reconstructs the original dataset from the same latent variables. This process forces the network to learn key features about the dataset, and it can be applied directly to our segmented OCT images (rather than the raw OCT data) to describe the structural signature of the glaucomatous ONH.

\begin{figure}[!ht]
	\centering
	\includegraphics[height = 8cm, width = 10cm]{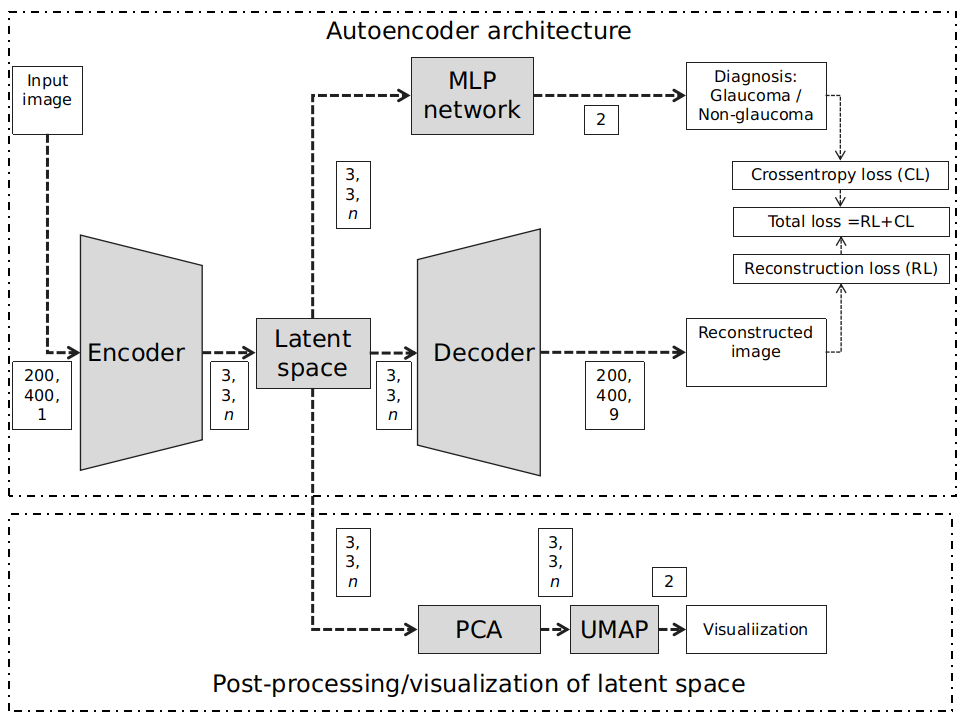}
	\caption{The architecture of the autoencoder network. The encoder reduced the dimension of an input image, whereas the decoder and the MLP branch reconstructed the original image and predicted the class of the image as either glaucoma or non-glaucoma, respectively. The PCA branch performed a coordinate transformation and reoriented the latent space along with the principal directions, whereas the UMAP block further reduced the latent space into a two-dimensional space for better visualization.  }
	\label{fig:Autoencoder}
\end{figure}

For our autoencoder, we used convolutional layers in the encoder portion to reduce the dimension of segmented OCT images into a low dimensional latent space (Fig.~\ref{fig:Autoencoder}, detailed architecture in the appendix). The decoder had a series of upsampling and deconvolution layers to reconstruct the original segmented image using the variables of the latent space. We also added a branch in parallel with the decoder for classification of images as either glaucoma or non-glaucoma (see Fig.~\ref{fig:Autoencoder}). This branch deployed a series of multilayer perceptron (MLP) layers followed by a softmax layer for classification. Therefore, the autoencoder network had access to the labels of each image during the training process. All three branches were trained simultaneously for 1,000 epochs, and the weights of the epoch with the lowest validation loss were considered as the best weights for the network. The loss function was the sum of the image reconstruction loss obtained from the decoder branch and the crossentropy loss obtained from the image classification branch (see appendix for more details).
Our training, validation, and test sets had 5,290 ($70.2\%$), 1,130 ($15\%$), and 1,111 ($14.8\%$) segmented OCT images, respectively. The split was performed in such a way that there would not be any duplicate images or images from the same subject in different sets. Furthermore, we balanced the training and the validation dataset (for different studies and classes) in such a way that each study would have the same number of non-glaucoma and glaucoma images, and all four studies would have the same number of images. 
The input OCT images were $200 \times 400$ pixels and were reduced to a low dimensional space by the encoder. To identify the optimal latent dimension that can adequately describe the high-dimensional segmented OCT images (of original dimension $200 \times 400 = 80,000$), we constructed eight different autoencoder models, each with a different latent-space dimension (ranging from 9 to 72 in steps of 9). Each autoencoder was trained separately. The performance of each model was then assessed by computing the Dice coefficient for the reconstructed images in the test dataset, and all models were compared for performance. We also estimated and compared the area under the receiver operating characteristic curves (AUCs), accuracies, and sensitivities (fixed at $95\%$ specificity) for all models (see appendix for more details). Five-fold cross-validation was performed to assess the effectiveness of each network.

\subsection{PCA and UMAP for dimension reduction and visualization}
\paragraph{}
Our goal was to understand how changing the value of a latent feature (output of the autoencoder) described a specific structural change of the ONH (e.g. change in RNFL thickness or disc radius). To this end, at first, we performed PCA on the latent features of the autoencoder and identified the principal directions. Subsequently, we altered the magnitude of each principal component (PC) in ten steps and reported how it impacted the morphology of the ONH tissues. This was done qualitatively by reconstructing segmented OCT images using the decoder in each step and then by comparing them with the baseline image.

To understand the importance of each PC, a machine learning classifier, i.e., support vector classifier (SVC), was used to evaluate the diagnostic efficiency of each PC; we ranked them in terms of their discriminating capabilities. This provided information about the PCs and their corresponding ONH structural features that are the most relevant for glaucoma detection.

The PCA is a linear projection method, and it is not efficient to capture the nonlinear trends in the data. Therefore, we used it only to make the latent variables untangled and not for dimension reduction. To facilitate visualization, we used the efficient and nonlinear Uniform Manifold Approximation and Projection (UMAP) technique to further reduce the latent dimensions to a 2D space \cite{mcinnes2018umapsoftware}. With this, we were able to describe each ONH by two numbers in the UMAP latent space, i.e., UMAP latent features 1 and 2. This post-processing method is shown in Fig.~\ref{fig:Autoencoder} as a separate module. UMAP allowed us to visualize ONH shapes in a 2D space and understand how they changed when transitioning from the normal to the glaucoma zones.

\section{Results}

\subsection{Segmentation network performance}
\begin{figure}[htp]
     \centering
     \begin{subfigure}[b]{0.65\textwidth}
         \centering
         \includegraphics[height = 6cm, width = 11cm]{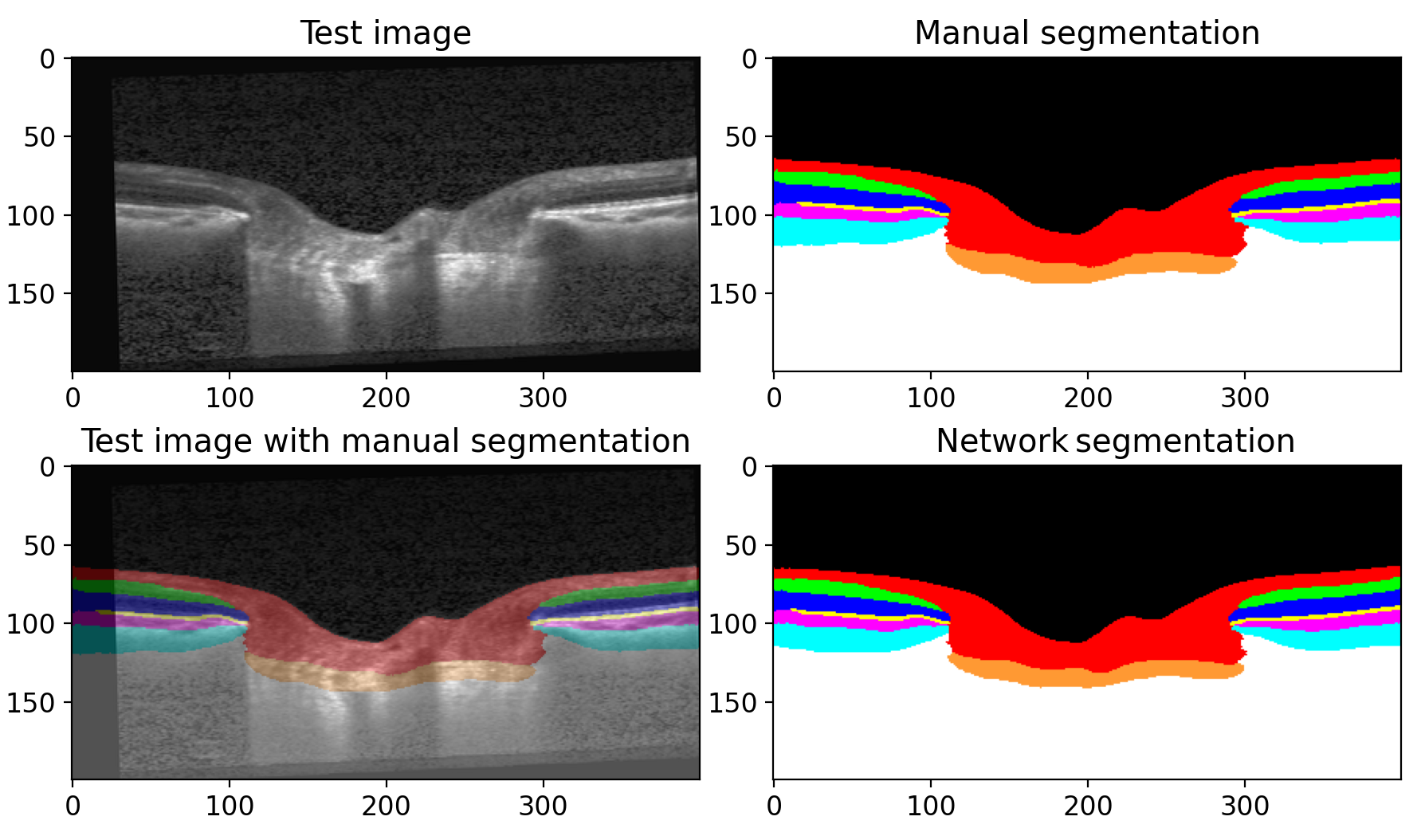}
         \caption{}
         \label{fig:SegmentationImage}
     \end{subfigure}
     \begin{subfigure}[b]{0.3\textwidth}
         \centering
         \includegraphics[height = 6cm, width = 3.5 cm]{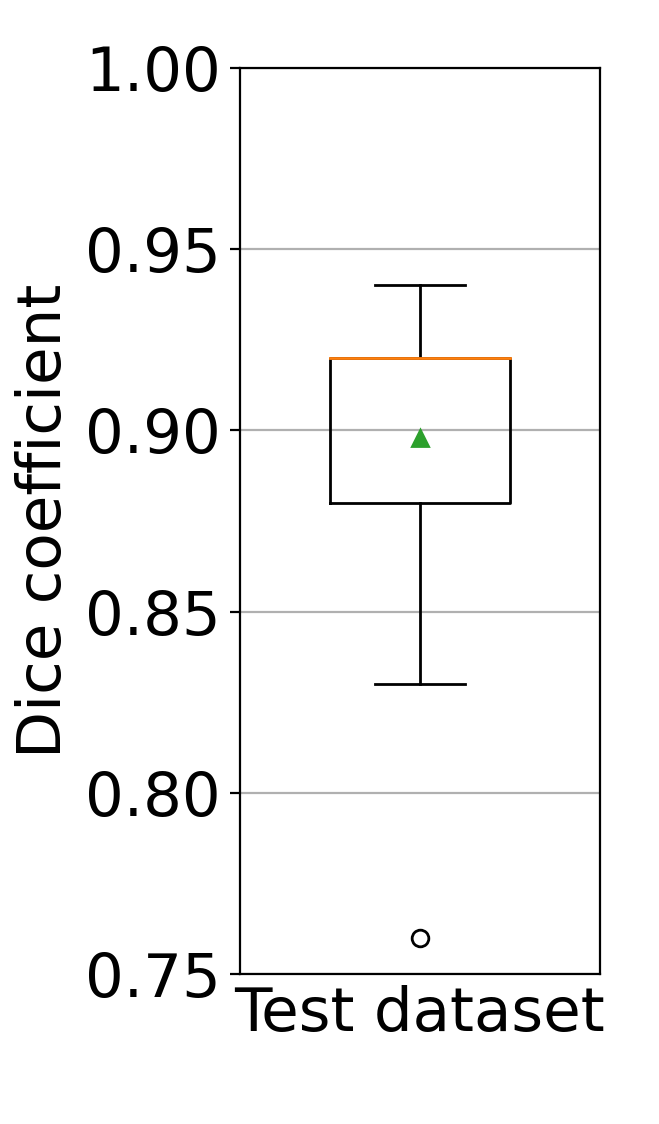}
         \caption{}
         \label{fig:SegmentationDiceCoefficient}
     \end{subfigure}
        \caption{Segmentation network performance. a) A test OCT image (top left) with manual (top right) and network generated (bottom right) segmentation images. The bottom left image shows the superimposed OCT and manually segmented images for comparison. b) The distribution of the Dice coefficient as a box plot. }
\end{figure}

\paragraph{}
Our segmentation network was able to delineate and isolate all neural and connective tissue layers of the ONH simultaneously. Fig.~\ref{fig:SegmentationImage} shows a sample OCT image (top left) from the test dataset with the corresponding manual segmentation (top right) and the network generated segmentation (bottom right).

The network-generated segmentations were found to be comparable and consistent with the manual segmentations. The quality of the segmentation was quantified by computing the Dice coefficient for the test images. Mean and standard deviation (SD) of the Dice coefficient were reported for the test set. Fig.~\ref{fig:SegmentationDiceCoefficient} shows the variation in Dice coefficients as a boxplot with mean and SD values as $0.90\pm 0.04$. It demonstrates the efficacy of the network for segmenting OCT images.

\subsection{Autoencoder network performance}
\paragraph{}
We noticed that the image reconstruction quality improved with the latent size. The average value of the Dice coefficient on 1,111 test images, with nine latent dimensions, was $0.75\pm 0.06$ (Table \ref{table:Autoencoder latent space size}). This value gradually increased with the number of latent variables up to the size of 54, and then remained relatively stable. The highest Dice coefficient achieved was $0.86\pm 0.04$ with 54 dimensions.  For the binary classification task, the values of AUC, accuracies, and sensitivities at $95\%$ specificity were found to increase with the size of the latent space; however, no significant change ($p<0.01$) was observed after 54 dimensions (see Table \ref{table:Autoencoder latent space size}). Therefore, the optimal size of the latent space was chosen to incorporate 54 dimensions. Table \ref{table:Autoencoder latent space size} shows the mean and SD values of the matrices used for comparison of models with different sizes of latent space.

\begin{center}
\begin{table}[htp]
\centering
\begin{tabular}{ |p{2cm}|p{2cm}|p{2cm}|p{2cm}|p{2cm}| } 
\hline
Latent space dimension & Dice coefficient & Accuracy ($\%$) & Sensitivity at $95\%$ specificity ($\%$) & AUC ($\%$) \\
\hline
9  & $0.75\pm 0.06$	& $88.3\pm 2.9$ & $86.2\pm 2.4$ & $87.5\pm 3.5$\\
18 & $0.78\pm 0.08$	& $88.4\pm 3.1$ & $86.5\pm 2.1$ & $89.3\pm 3.2$\\
27 & $0.80\pm 0.06$	& $89.1\pm 3.0$ & $86.9\pm 3.1$ & $88.2\pm 3.8$\\
36 & $0.81\pm 0.05$	& $90.3\pm 2.5$ & $87.4\pm 3.4$ & $90.0\pm 3.2$\\
45 & $0.84\pm 0.04$	& $90.7\pm 2.8$ & $89.6\pm 3.1$ & $91.2\pm 2.8$\\
54 & $\textbf{0.86}\pm 0.04$	& $91.8\pm 2.4$ & $\textbf{90.0}\pm 2.4$ & $93.3\pm 3.2$\\
63 & $0.86\pm 0.05$	& $\textbf{92.0}\pm 2.3$ & $89.9\pm 2.8$ & $93.1\pm 2.9$\\
72 & $0.86\pm 0.06$	& $91.7\pm 2.5$ & $89.8\pm 2.7$ & $\textbf{94.1}\pm 2.0$\\
\hline
\end{tabular}
\caption{Performance comparison of autoencoders with different latent dimensions.}
\label{table:Autoencoder latent space size}
\end{table}
\end{center}

\begin{figure}
     \centering
     \begin{subfigure}[b]{0.45\textwidth}
         \centering
         \includegraphics[height = 7cm, width = 6cm]{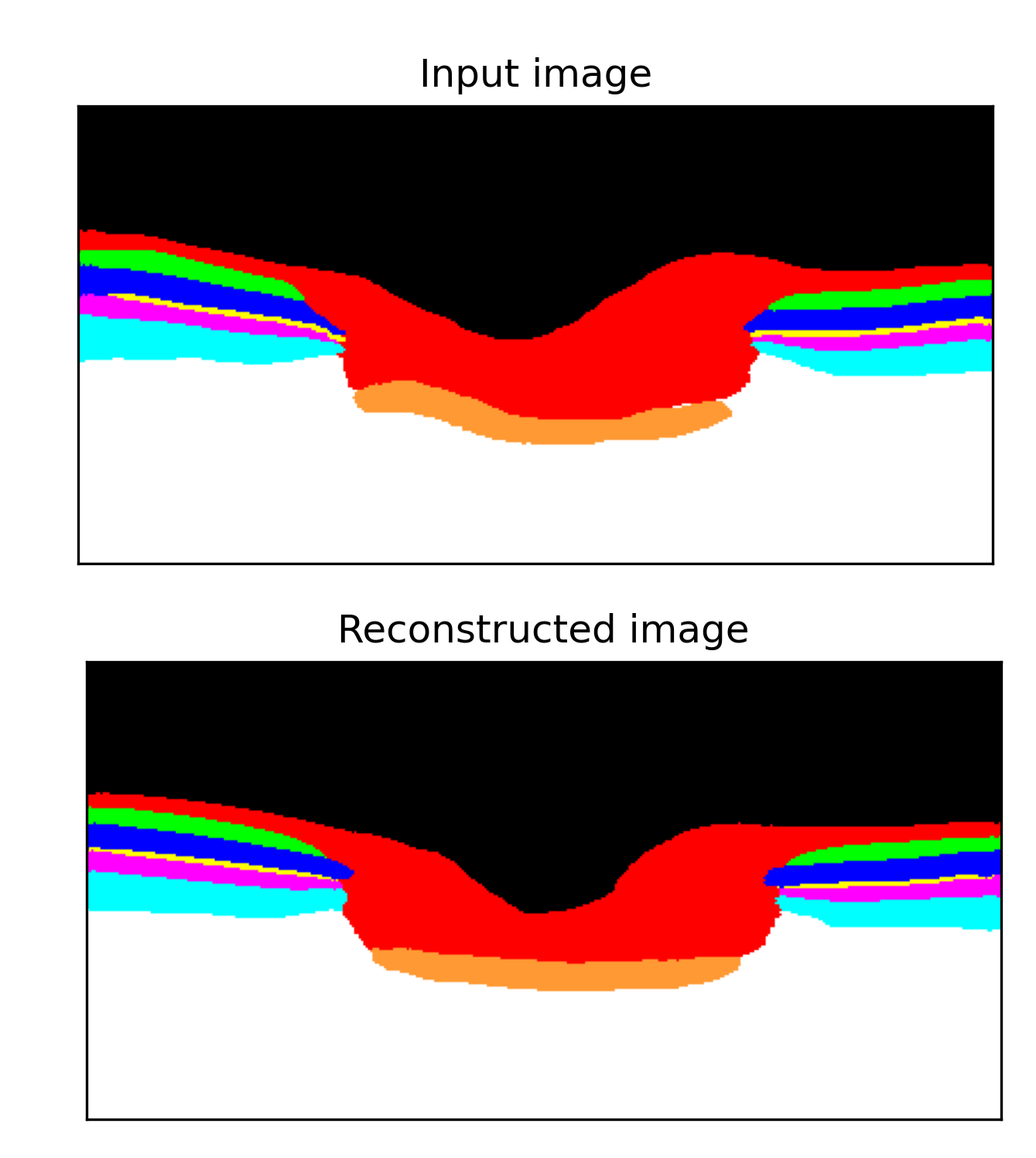}
         \caption{}
         \label{fig:AutoencoderImage}
     \end{subfigure}
     \begin{subfigure}[b]{0.45\textwidth}
         \centering
         \includegraphics[height = 7cm, width = 4 cm]{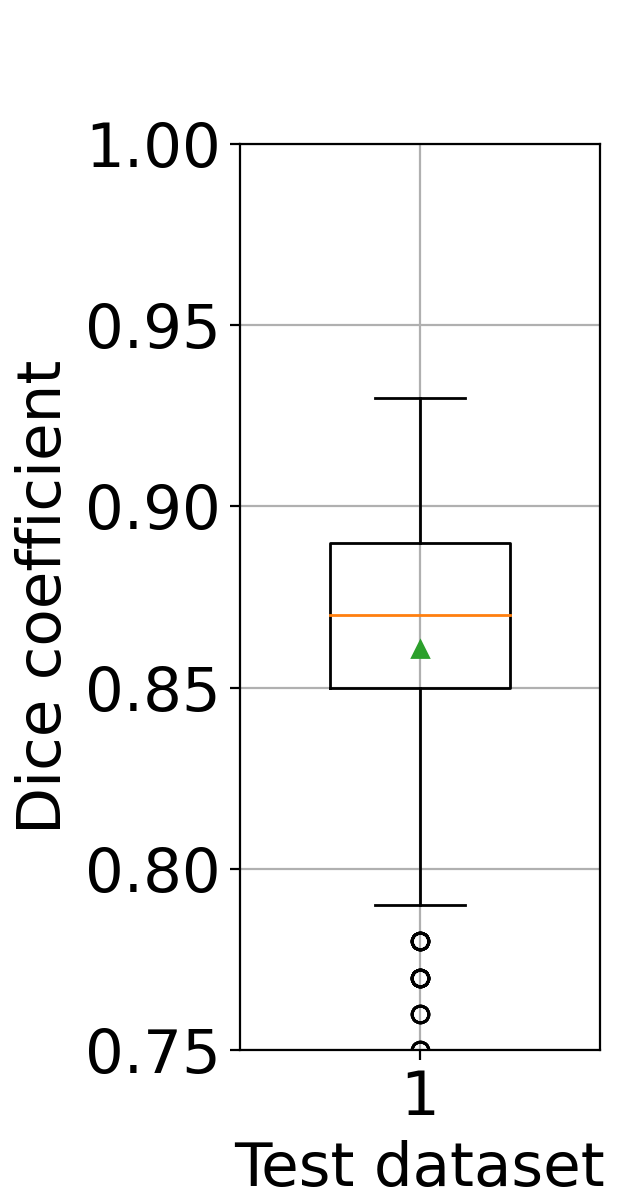}
         \caption{}
         \label{fig:AutoencoderDiceCoefficient}
     \end{subfigure} \\
     \begin{subfigure}[b]{0.45\textwidth}
         \centering
         \includegraphics[height = 6cm, width = 6 cm]{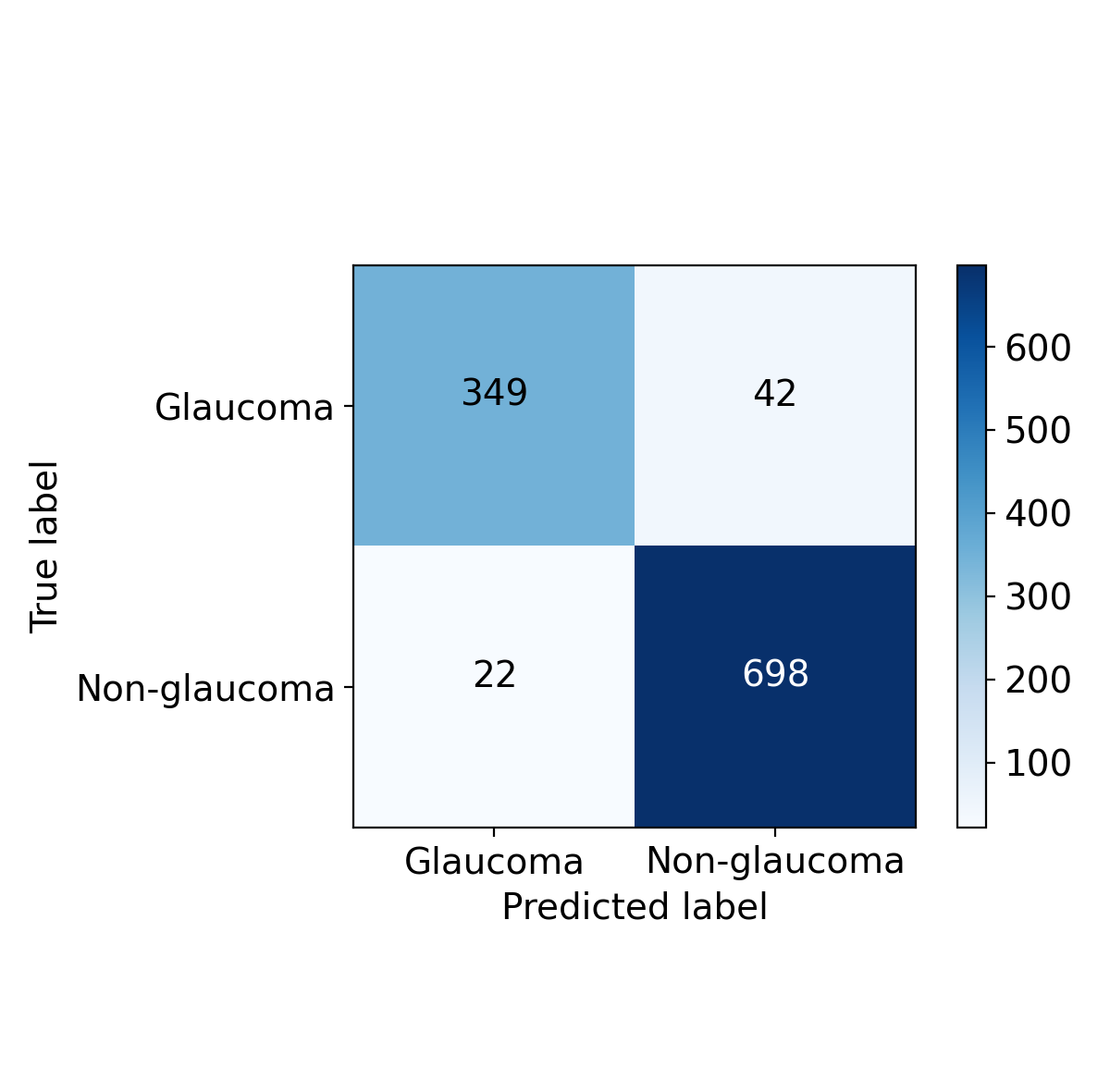}
         \caption{}
         \label{fig:ConfusionMatrix}
     \end{subfigure}
     \begin{subfigure}[b]{0.45\textwidth}
         \centering
         \includegraphics[height = 6cm, width = 6 cm]{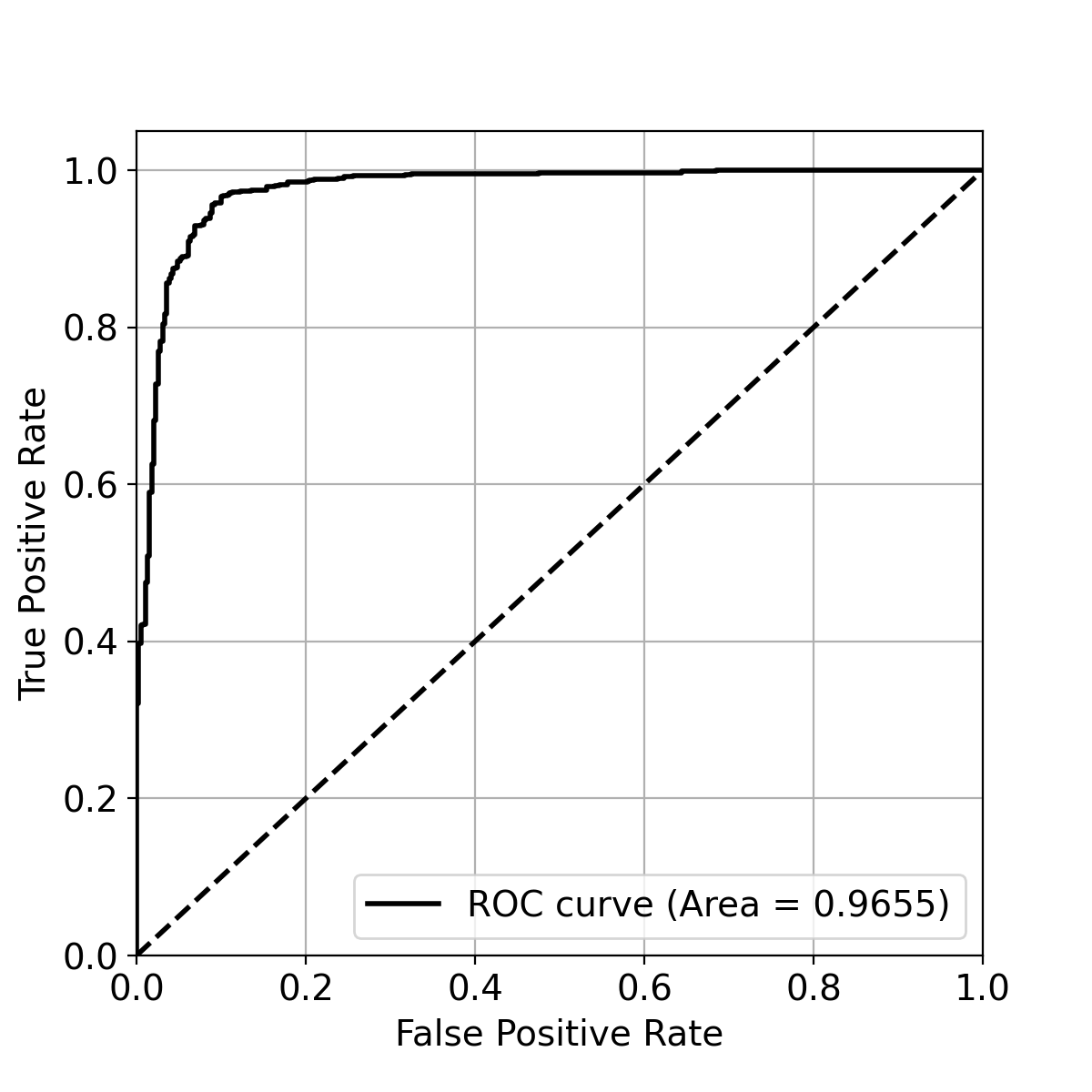}
         \caption{}
         \label{fig:ROC}
     \end{subfigure}
     \label{fig:Autoencoder network performance}
        \caption{Network performance with 54 latent dimensions. a) An input image from the test dataset (top) and the corresponding reconstructed image by the network (bottom). b) The distribution of the Dice coefficients with 1,111 test images as a box plot. c) The confusion matrix for the binary classification. d) Receiver operating characteristic curve for the classification task.}
\end{figure}

Our autoencoder network with 54 latent dimensions was able to reconstruct the test images with good accuracy. Fig.~\ref{fig:AutoencoderImage} depicts a sample test image (top) with the corresponding reconstructed image (bottom). The average value of the Dice coefficient with 1,111 test images was $0.86\pm 0.04$ (Fig.~\ref{fig:AutoencoderDiceCoefficient}). We also achieved good diagnostic performance (Fig.~\ref{fig:ConfusionMatrix}) with $94\%$ accuracy ($1047/1111$ images were correctly classified),  $89\%$ sensitivity ($349/391$ glaucoma images were correctly classified), and $97\%$ specificity ($698/720$ healthy images were correctly classified). The highest value of AUC obtained with the test images was $0.96$ (Fig.~\ref{fig:ROC}), and at this threshold level, the sensitivity (for a fixed specificity of $95\%$)  was $92\%$.

\begin{figure}
     \centering
     \begin{subfigure}[b]{0.3\textwidth}
         \centering
         \includegraphics[height = 6cm, width = 3cm]{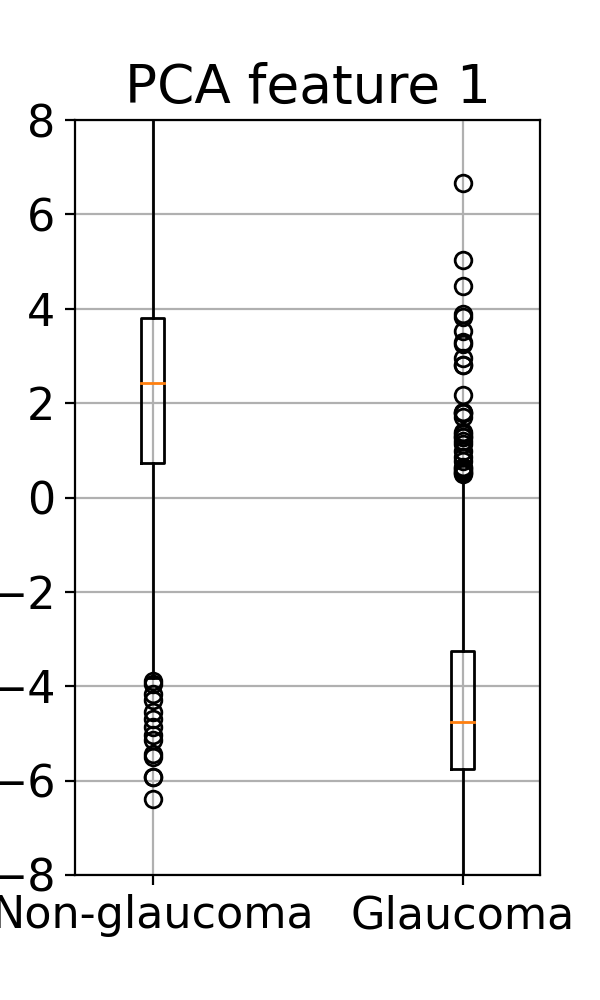}
         \caption{}
         \label{fig:PCA_pvalue_54_1}
     \end{subfigure}
     \begin{subfigure}[b]{0.3\textwidth}
         \centering
         \includegraphics[height = 6cm, width = 3 cm]{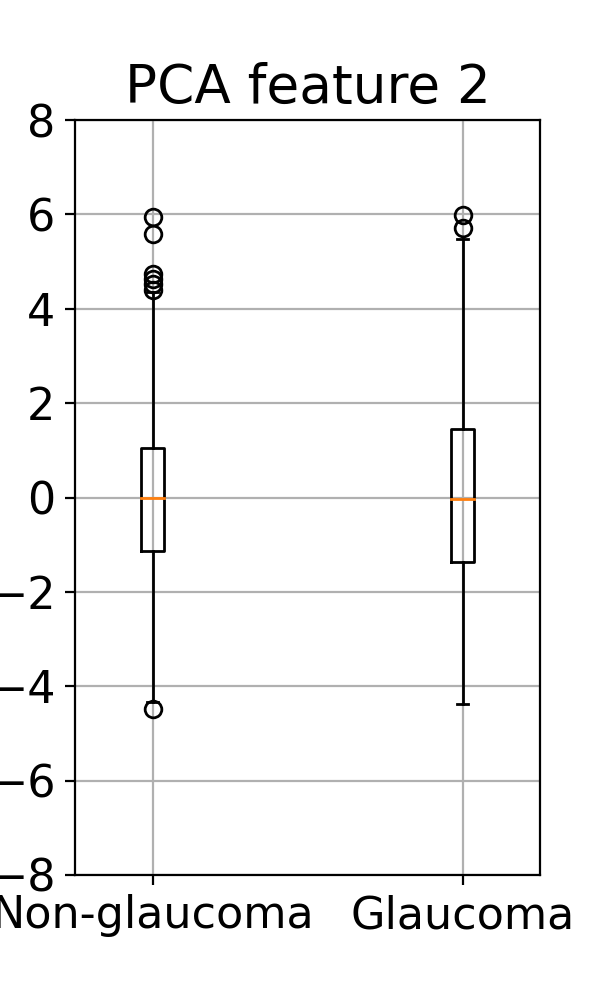}
         \caption{}
         \label{fig:PCA_pvalue_54_2}
     \end{subfigure}
     \begin{subfigure}[b]{0.3\textwidth}
         \centering
         \includegraphics[height = 6cm, width = 3 cm]{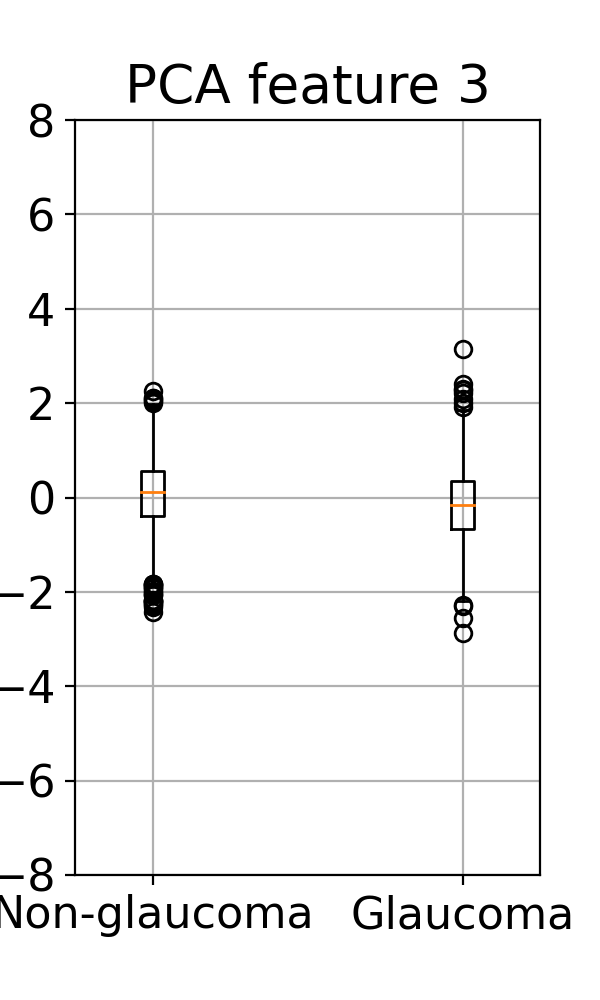}
         \caption{}
         \label{fig:PCA_pvalue_54_3}
     \end{subfigure} \\
     \begin{subfigure}[b]{0.45\textwidth}
         \centering
         \includegraphics[height = 6cm, width = 8 cm]{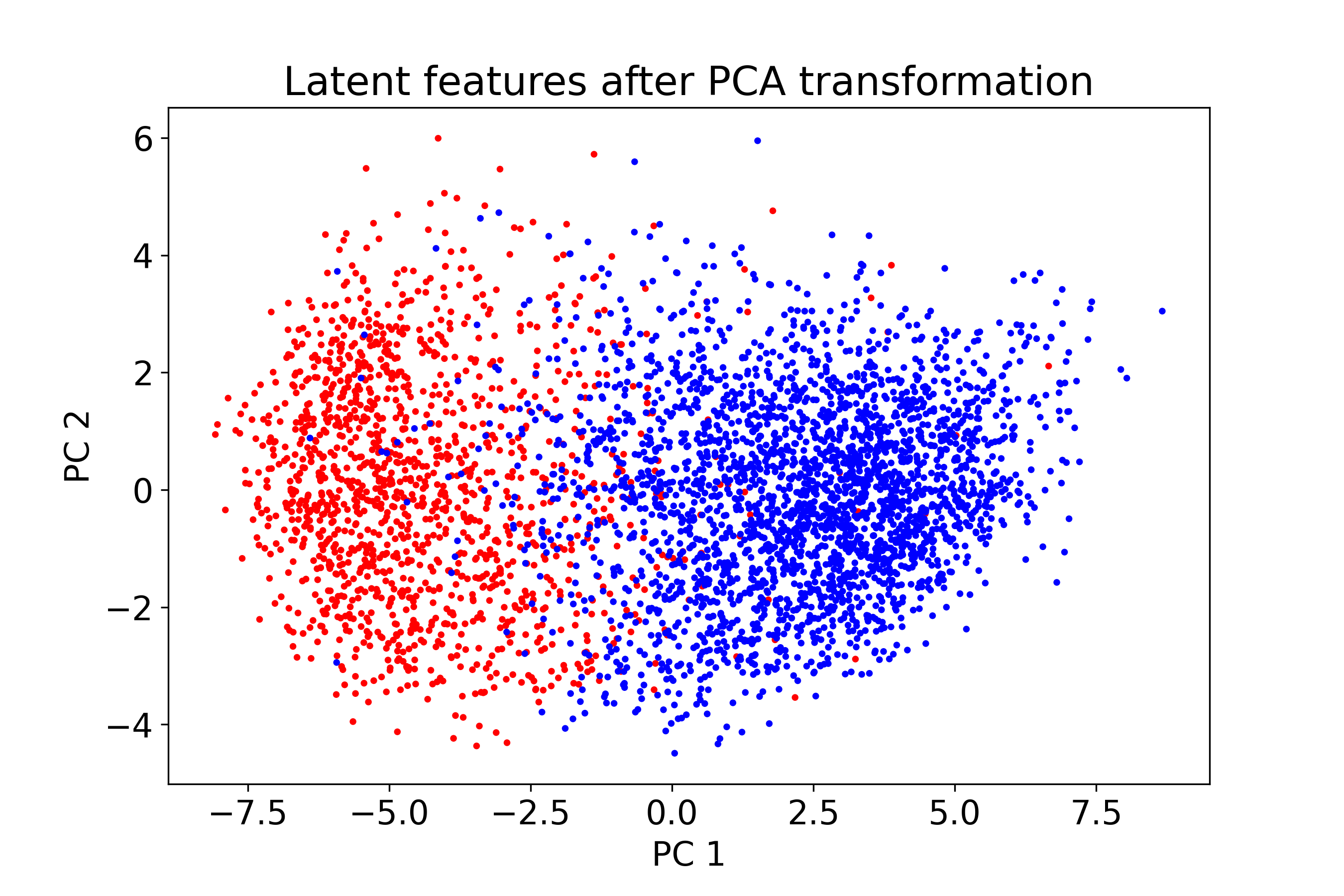}
         \caption{}
         \label{fig:PCA_clustering_54}
     \end{subfigure}
     \begin{subfigure}[b]{0.45\textwidth}
         \centering
         \includegraphics[height = 6cm, width = 8 cm]{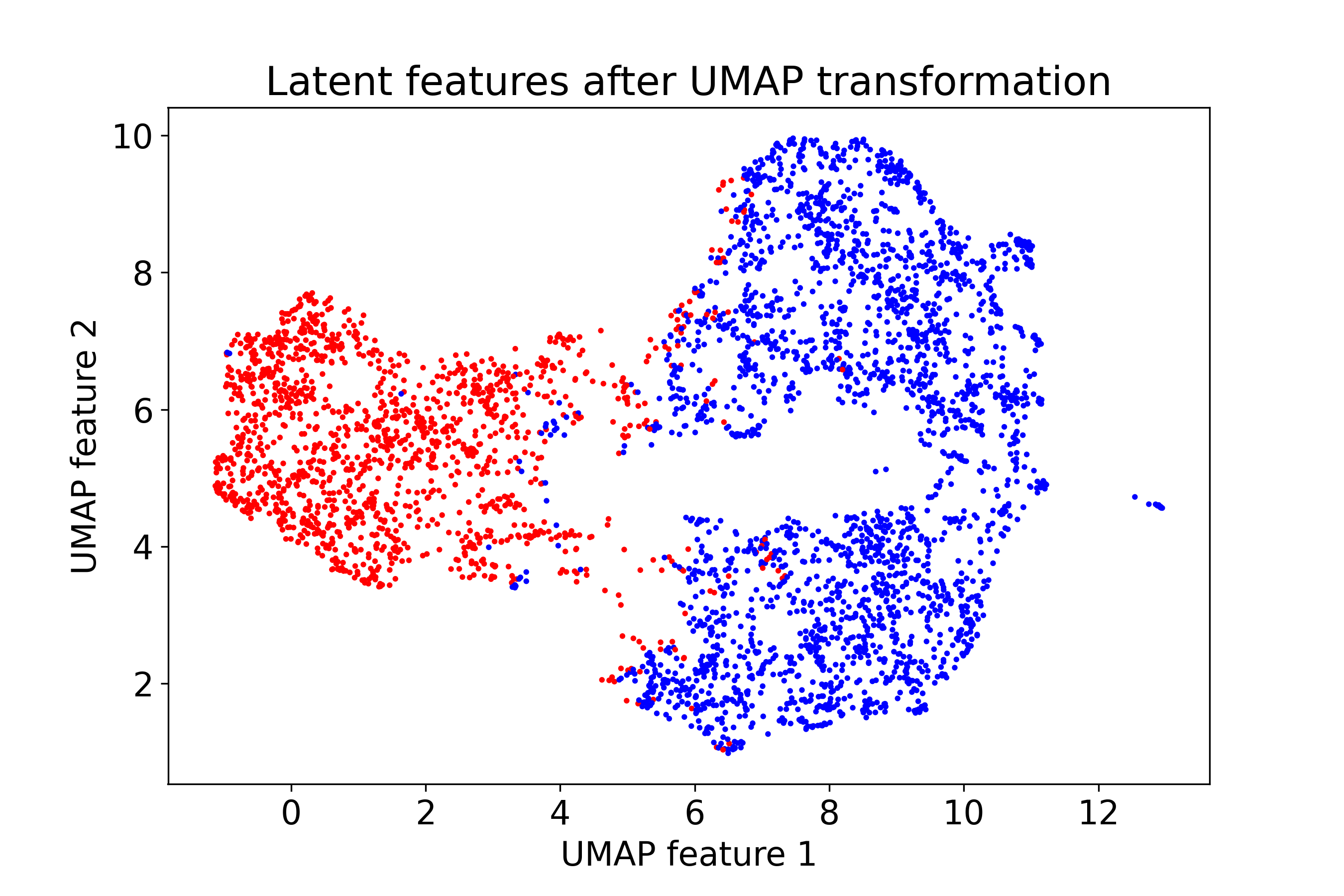}
         \caption{}
         \label{fig:UMAP_clustering_54}
     \end{subfigure}
        \caption{a-c) Box plots showing the distribution of PC1, PC2, and PC3 features for non-glaucoma and glaucoma eyes. d) Visualization of PC1 and PC2 features as a scatter plot after PCA transformation. Blue points: non-glaucoma, and red points: glaucoma e) Visualization of UMAP features as a scatter plot after UMAP transformation to reduce the transformed latent space size from 54 to 2.  }
\end{figure}

\subsection{PCA and dimension reduction with UMAP }
\paragraph{}
We performed PCA on the 54 latent dimensions to identify the 54 principal directions and then compared their values for the non-glaucoma and glaucoma eyes. It was found that the PC1 values for non-glaucoma and glaucoma eyes were significantly different ($p < 10^{-9}$), whereas the difference was relatively less for other PCs. Figs.~\ref{fig:PCA_pvalue_54_1}-\ref{fig:PCA_pvalue_54_3} show the distributions of PC1 ($p < 10^{-9}$), PC2 ($p<0.1$), and PC3 ($p < 10^{-6}$) for non-glaucoma and glaucoma eyes as box plots. In particular, the PC1 was the most influential and provided a diagnostic accuracy of $86.1 \pm 2.6 \%$. The addition of other PCs improved the accuracy slightly (see the cumulative accuracy column in Table \ref{table:Accuracies with PCs}). After the first five PCs, no significant differences in the values for non-glaucoma and glaucoma eyes were observed for other PCs.

\begin{center}
\begin{table}[h!]
\center
\begin{tabular}{ |c|c|c| } 
\hline
PC & Accuracy ($\%$) & Cumulative accuracy ($\%$) \\
\hline
PC1 & 86.1 $\pm$ 2.6 & 86.1 $\pm$ 2.6\\
PC2 & 69.1 $\pm$ 2.1 & 87.0 $\pm$ 1.5 \\
PC3 & 71.1 $\pm$ 1.4 & 88.3 $\pm$ 2.0 \\
PC4 & 61.1 $\pm$ 1.9 & 88.9 $\pm$ 1.3 \\
PC5 & 60.1 $\pm$ 2.0 & 89.1 $\pm$ 1.7\\
All &  & 91.8 $\pm$ 2.4\\
\hline
\end{tabular}
\caption{Diagnostic accuracy with individual PCs}
\label{table:Accuracies with PCs}
\end{table}
\end{center}

Using PCA and UMAP, we found that it was possible to dissociate non-glaucoma eyes from glaucoma eyes into two separate and distinct clusters by solely interpreting structural information of the ONH and without using any other clinical or demographic parameters (Fig.~\ref{fig:PCA_clustering_54} - \ref{fig:UMAP_clustering_54}, non-glaucoma eyes represented in blue and glaucoma eyes in red). UMAP performed better as compared to PCA for finding correlations within the image data and clustering them as it can recover the nonlinear relationships between the data points.

\subsection{Structural signature of the glaucomatous ONH }
\begin{figure*}[!ht]
	\centering
	\includegraphics[height = 12cm, width = 0.99\textwidth]{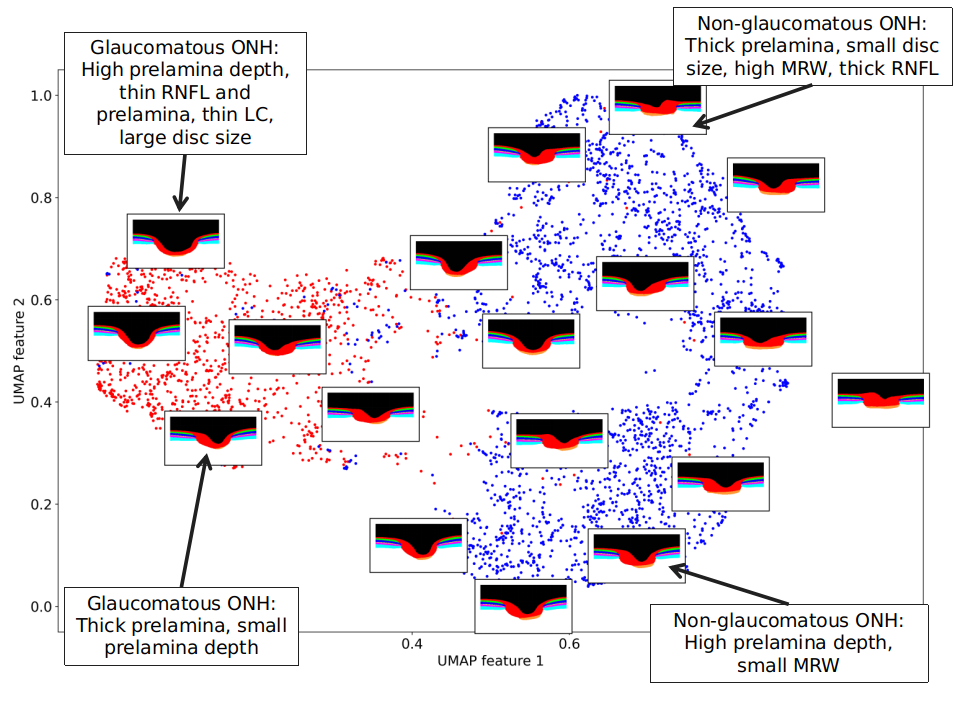}
	\caption{A few reconstructed segmented OCT images for non-glaucoma/glaucoma eyes in the 2D latent space of UMAP. }
	\label{fig:UMAP_OverlapReconImages}
\end{figure*}

\paragraph{}
To understand variations in ONH structural features between non-glaucoma and glaucoma eyes, we reconstructed a few ONHs in the UMAP space using the two UMAP latent features (Fig.~\ref{fig:UMAP_OverlapReconImages}). We found that the top-right region of the UMAP space described non-glaucoma ONHs with thick prelamina, small disc size, high MRW, and thick RNFL; whereas the non-glaucomatous ONHs with higher prelamina and smaller MRW were clustered in the bottom-right region. The central space represented the transition from non-glaucoma to glaucoma, where both glaucoma and non-glaucoma ONHs were intermixed. We observed that the glaucomatous ONHs with high prelamina-depth, thin prelamina, thin RNFL, thin LC, and large disc size were clustered together in the top-left region, whereas towards the bottom-left space, we observed glaucomatous ONHs with relatively thicker prelamina and smaller cup-depth.

\begin{figure*}[!ht]
	\centering
	\includegraphics[height = 4cm, width = 0.9\textwidth]{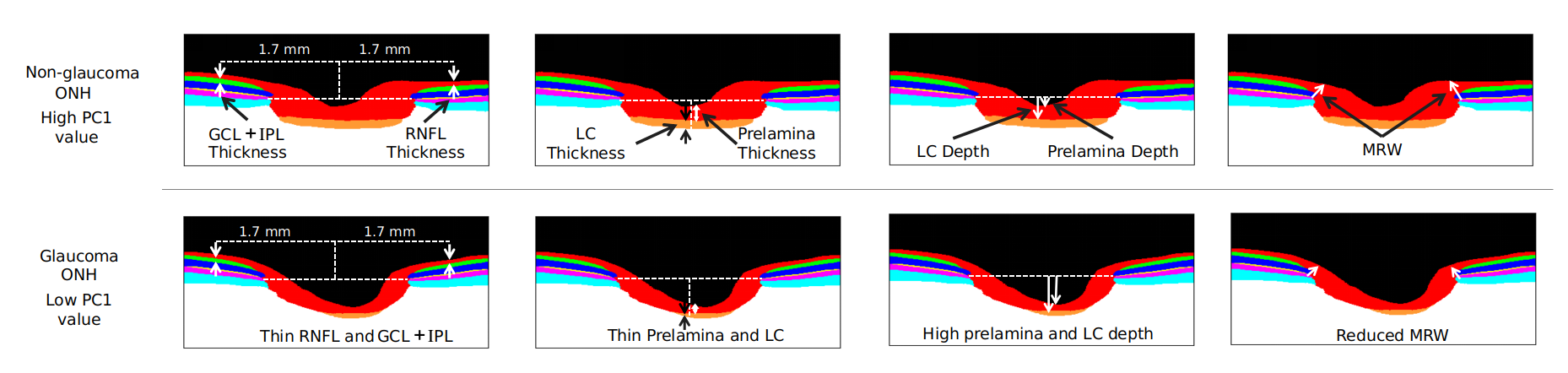}
	\caption{ONH Structural changes with variations in PC1 from a high (non-glaucoma eye) to a low (glaucoma eye) value. The images at the top represent a sample test image that was classified as non-glaucoma by the MLP network (high PC1 value), and those at the bottom are the glaucoma images (low PC1 value) obtained after reducing the value of PC1 while keeping all other PCs constant. Each column represents a key structural change while changing PC1. }
	\label{fig:PC1_Variation}
\end{figure*}

\begin{figure}
     \centering
     \begin{subfigure}[b]{0.3\textwidth}
         \centering
         \includegraphics[height = 6cm, width = 2.5cm]{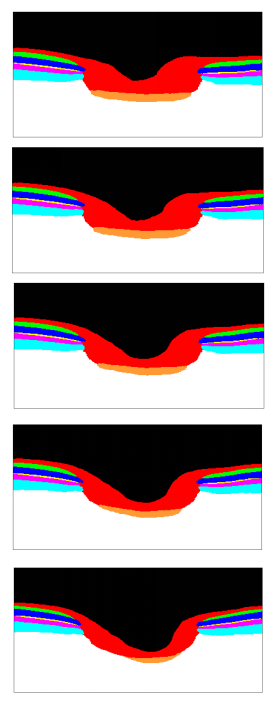}
         \caption{}
         \label{fig:journey_discrete}
     \end{subfigure}
     \begin{subfigure}[b]{0.6\textwidth}
         \centering
         \includegraphics[height = 6cm, width = 11 cm]{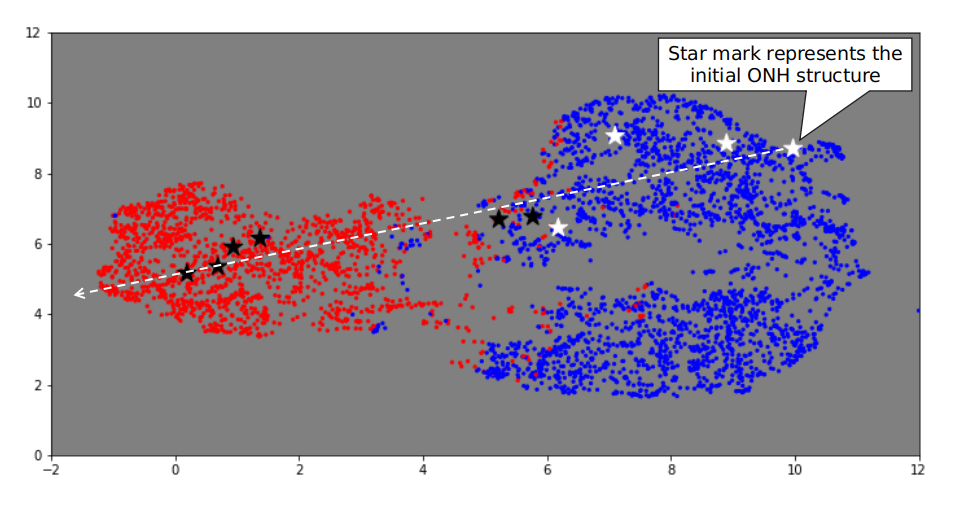}
         \caption{}
         \label{fig:journey}
     \end{subfigure}
        \caption{ONH shape transition from the non-glaucoma (blue point cloud) to the glaucoma (red point cloud) zone while varying PC1 for a given non-glaucoma ONH starting point. PC1 was changed in 10 steps (shown as 10 stars in the UMAP latent space) and five representative shapes are shown in the left column in (a), for steps: 1, 4, 6, 8, and 10. All ONH shapes were classified by the MLP network (white stars: non-glaucoma ONHs; black starts: glaucoma ONHs). The top-right, white star mark in (b) shows the position of the initial ONH in the UMAP latent space. The white dashed line shows the direction for PC1 changes (high to low value).  }
        \label{fig:journey_all}
\end{figure}

To understand the influence of PCs on the shape of the ONH, each can be varied one-at-a-time, starting with PC1- the component that contributed the most to glaucoma diagnosis. As PC1 varied from a high value (non-glaucoma eye) to a low value (glaucoma eye), we generally observed thinning of the RNFL, GCL+IPL, prelamina, and LC layers, which was accompanied with a decrease in MRW. The depth of the prelamina and LC also increased. These key structural changes can be observed in Fig.~\ref{fig:PC1_Variation} starting from a sample non-glaucoma ONH. When the change in PC1 was further divided into 10 steps (high to low value), we observed how a non-glaucoma (blue point cloud) ONH transitioned into the glaucoma zone (red point cloud) as shown in Fig.~\ref{fig:journey_all}.

We also found that changes in PC2 increased the prelamina depth, LC depth, disc size, and reduced the prelamina thickness. The ONH tissues were also found to bow posteriorly. However, no changes in RNFL, GCL+IPL, and LC thickness were observed (more details in the appendix). The PC3 represented an increase in disc size, thinning of the prelamina, and a decrease in MRW; however, we did not witness any changes in prelamina/LC depth and layer thicknesses except for the prelamina. The outcomes of changing other PCs are presented in the appendix, and schematics representing ONH structural changes with variations in PCs are given in the supplementary material.

The first three PCs were the most important to describe ONH structural changes, and with changes in their magnitudes, ONHs made transitions from the non-glaucoma to the glaucoma zone. On the other hand, the contributions of other PCs were not significant for discriminating between non-glaucoma and glaucoma ONHs, and they are more likely to represent inter-individual variations in the non-glaucoma group.

\section{Discussion}
\paragraph{}
An autoencoder was applied to a large OCT dataset to identify novel features of ONH. The autoencoder was customized to have an MLP branch in parallel with the decoder network so that image reconstruction and classification can be performed simultaneously. The network identified objective and quantitative structural measurements of ONH that were crucial for glaucoma diagnosis. We also used a post-processing/visualization module (see Fig.~\ref{fig:Autoencoder}) to reduce the multi-dimensional latent space to 2D through a series of transformations. The dimension reduction method demonstrated that it was possible to cluster non-glaucoma and glaucomatous eyes in two separate groups in a low dimensional latent space. It was also found that each PC can be correlated to a few structural features of the ONH, and then, by perturbing latent representations in the principal directions and using the UMAP latent space, we were able to reveal the ONH structural changes that occur while making a transition from the non-glaucoma to the glaucoma zone (Fig.~\ref{fig:journey_all}).

In this study, we found that the novel biomarkers, extracted by our autoencoder network form segmented OCT images, could classify glaucoma and non-glaucoma eyes with high accuracy (diagnostic accuracy: $94\%$ and AUC: $0.96$). In the literature, several studies have reported glaucoma diagnosis using different techniques, e.g., the use MLCs with demographic or morphometric parameters \cite{kim2009retinal, huang2005development, mwanza2011ability, silva2013sensitivity}, the use of supervised or unsupervised AI networks with visual field maps \cite{asaoka2016detecting, kim2017development, yousefi2014learning}, color fundus images \cite{berchuck2019estimating, diaz2019retinal, gulshan2016development}, RNFL maps \cite{kim2017development,barella2013glaucoma}, and 2D or 3D raw OCT scans \cite{maetschke2019feature, petersen2020data}. To the best of our knowledge, no studies have yet reported a glaucoma diagnosis solely from segmented OCT images. In this work, we proposed a novel method for diagnosis that utilized only one segmented OCT image from each subject. Without using any additional information, e.g., demographic parameters or information about texture, edge, and intensity, our network solely focused on the morphology of the layers and their spatial relations to differentiate glaucoma from non-glaucoma eyes. We provided information about neural (e.g., RNFL, GCC, and GCL+IPL) and connective tissues (e.g., choroid, sclera, and LC) of the ONH to the network so that the morphological features of all the layers could be exploited for diagnosis. Furthermore, we did not force the autoencoder to learn any human-defined biomarkers; however, it discovered a set of parameters that were closely related to clinical observations (see Fig.~\ref{fig:PC1_Variation}).

We observed that the first three PCs (out of 54) solely contributed to glaucoma diagnosis. PC1 explained the simultaneous variations in the thickness of different layers (e.g., thinning of the RNFL, GCL+IPL, prelamina, and LC layers) during glaucoma development. PC1 alone differentiated non-glaucoma from glaucoma eyes very accurately (see Fig.~\ref{fig:PCA_pvalue_54_1}) and provided a diagnostic accuracy of $86.1 \pm 2.6\%$. It also explained the reason behind the high diagnostic accuracy even at low latent dimensions (see Table \ref{table:Autoencoder latent space size}). The diagnostic accuracy with nine latent dimensions was $88.3 \pm 2.9 \%$, and it increased by only $4\%$ with a seven-fold increase in latent dimensions (see Table \ref{table:Autoencoder latent space size}). One of the most widely accepted standards for glaucoma diagnosis/prognosis is the analysis of the RNFL thickness \cite{mwanza2011ability, barella2013glaucoma, an2018comparison}. A few studies have also advocated the benefits of the ganglion cell complex thickness \cite{miki2015retinal, kita2013ability}. However, these measurements may not always be suitable, as in a few studies, they were found to be stable after reaching a minimum threshold, even in the case of severe glaucoma \cite{pavlidis2003retinal}. RNFL and GCC thickness measurements can also be drastically affected by blood vessels from the central retinal vessel trunk - itself providing a skeleton to maintain the shapes of neural and connective tissues even in the presence of damage \cite{zhang2018central}. Since PC1 summarized multiple structural variations into a single number, it may prove to be superior to the measurement of RNFL or GCC thickness alone and could be considered for glaucoma monitoring.

In this study, we found that PC2 and PC3 described the change in disc size and bowing of the peripapillary sclera. Both have been described in the literature as potentially critical parameters to better understand glaucoma progression \cite{hancox1999optic,hoffmann2007optic, quigley2015contribution}. But no studies have yet compared and ranked the diagnostic ability of these parameters with that of retinal cell layer thicknesses. We observed that the PC1 features were more significant than PC2-3 and ranked them in terms of their diagnostic abilities (Table \ref{table:Accuracies with PCs}).

Overall, PC1, PC2, and PC3 represented the thickness of neural layers, prelamina/LC thickness and depth, ONH curvature, MRW thickness, and disc size; and also revealed the structural changes that are most likely to appear in the ONH during glaucoma development. Although the measurement of RNFL and GCC thickness is a widely accepted criteria for glaucoma diagnosis, these hand-engineered parameters may not be sufficient to define the complex structural changes that appear in the ONH during glaucoma development. Since our autoencoder recommended PCs summarized simultaneous changes in ONH, they may act as novel and new biomarkers for diagnosis.

Currently, a popular trend for diagnosis is the use of CNN networks. These deep-learning-based approaches have proved their diagnostic efficacy with high AUC values \cite{maetschke2019feature, gulshan2016development, petersen2020data, muhammad2017hybrid}. These methods can also highlight the regions-of-importance of the ONH tissue through class activation maps (CAM) that are useful for diagnosis. However, these maps largely depend on the network architecture, and different architecture may lead to different regions-of-importance \cite{maetschke2019feature}. Furthermore, CAM only provides qualitative information. In this context, our method provided a novel approach that can extract meaningful information about the features used for diagnosis.

The other PCs (PC4-54) did not have a major influence in classifying non-glaucoma from glaucoma eyes (diagnostic accuracy increased from $88.3 \pm 2.0 \%$ to $91.8 \pm 2.4\%$ with the addition of all these PCs to the first three PCs, Table \ref{table:Accuracies with PCs}); however, they were beneficial for image reconstruction, as they increased the value of the Dice coefficient. These PCs may relate to intra-individual variations in the normal population, age-related structural changes, and they may also be linked to other conditions, such as myopia, which we have not excluded from this study. Our future work will investigate these trends in more details, and our goal is to fully understand the structural phenotype of the glaucomatous ONH.

In this study, UMAP provided an elegant way to visualize glaucoma and non-glaucoma ONHs in a 2D space. It classified glaucoma and non-glaucoma ONHs in two separate groups. In each group, ONHs were again clustered in terms of features, e.g., ONHs with thick prelamina, small disc size, and thick RNFL were gathered in the top-right corner of the non-glaucoma group (Fig.~\ref{fig:UMAP_OverlapReconImages}). We observed that the first three PCs had a significant effect on classification, and by changing their magnitude, it was possible to make a transition from the non-glaucoma zone to the glaucoma zone in the UMAP space (Fig.~\ref{fig:journey_all}). Because morphological changes of the ONH are often difficult to interpret for clinicians, our approach may provide a way to monitor these changes and visualize how a patient may progress into glaucoma over time. In the future, our network will be enhanced to additionally predict changes in visual field loss progression which could also improve our ONH structural classifications.

Unsupervised learning is an efficient way to learn novel biomarkers in a dataset. In recent years, many studies have used autoencoders, variational autoencoders, and PCA for this task \cite{kim2017development, yousefi2014learning, higgins2016beta}. Autoencoders are preferable for learning latent features as they can reveal complex nonlinear relations, whereas PCA is a linear transformation \cite{sakurada2014anomaly}. With some initial trials, we found that although unsupervised autoencoders were able to reconstruct segmented OCT images with high accuracy, the features extracted by them were inefficient for diagnosis, and subsequent PCA and UMAP transformations were not able to cluster the non-glaucoma and glaucomatous eyes separately (see appendix for more details). Therefore, we used a hybrid technique to exploit the advantages of both unsupervised and supervised training. While training the decoder network for image reconstruction, the MLP network was trained simultaneously for the classification task.

In this study, several limitations warrant further discussion. First, we used OCT images obtained from a single device, i.e., Spectralis. Since the quality of OCT images is device-dependent, segmentation networks trained with data from one machine may not perform well on data from others \cite{devalla2018drunet}. A few recent studies have already demonstrated device-independent and label-free techniques for OCT image segmentation \cite{DevallaTowards, devalla2018device}. We aim to test our methodology with data from multiple devices in the near future. Second, we only used a single OCT B-Scan (central horizontal B-Scan), from each OCT volume of each subject to train our network. Including multiple scans (or radial scans) might be helpful to learn other novel structural biomarkers about the ONH. Third, for the glaucoma group, we did not include information about glaucoma type, severity, and visual field loss, and for the non-glaucoma group, we did not include any other demographic/clinical information such as age or refractive error. This information will be considered in the next generation of our network, as it may help us understand, e.g., how PCs are also linked to visual field loss progression.

In summary, this study was able to demonstrate the potential use of autoencoders to reveal clinically significant, novel structural biomarkers and ONH signatures. The network was able to show the structural changes that appear in the ONH during glaucoma development and provided high diagnostic accuracy from a single segmented OCT B-Scan, which has not been proposed so far. The paradigm introduced in this study has many clinical applications and may help increase diagnosis accuracy, predict future patterns, and improve prognosis.

\begin{small}

\section*{Acknowledgments}
\paragraph{}
This work was supported by the Singapore Ministry of Education Academic Research Funds Tier 1 (R-397-000-294-114 (MJAG)), the Singapore Ministry of Education Tier 2 (R-397-000-280-112, R-397-000-308-112 (MJAG)), and the National Medical Research Council (Grant NMRC/ STAR/ 0023/ 2014 (TA)).

\bibliographystyle{ieeetr}
\bibliography{autoencoder.bib}

\newpage
\appendix

\section{U-net architecture}
\paragraph{}
The customized U-net had the same architecture as that of the original one proposed by Olaf Ronneberger et al. \cite{ronneberger2015u}. However, it had different number of filters in each level. The input to the network was a grayscale OCT image of size $200\times400$ pixels. The network was composed of a  contraction path (an encoder) and a symmetric expanding path (a decoder, see Fig.\ref{fig:UNet}). The layers of the network were as follows:

\begin{enumerate}
\item 
Conv1: Two 2D convolution (Conv2D) layers with 8 filters (stride = (3,3), padding = same,  kernel initializer = `he normal'), followed by a 2d max-pooling layer (MaxPooling2D) with pool size=(2, 2).
\item
Conv2: Same as Conv1 with 16 filters.
\item
Conv3: Same as Conv1 with 32 filters.
\item
Conv4: Same as Conv1 with 64 filters.
\item
Drop4: A dropout layer with rate = 0.5.
\item
Conv5: Same as Conv1 with 64 filters.
\item
Drop5: A dropout layer with rate = 0.5.
\item
UP6: UpSampling2D with size = (1,1) followed by a Conv2D layer with 64 filters (stride = (2,2), padding = same,  kernel initializer = `he normal'). A concatenate layer for concatenating with Drop4 along axis = 3
\item
Conv6: Two Conv2D layers with 64 filters
\item
Up7: UpSampling2D with size = (2,2) followed by a Conv2D layer with 32 filters. A concatenate layer for concatenating with Conv3 along axis = 3
\item
Conv7: Same as Conv6 with 32 filters.
\item
Up8: UpSampling2D with size = (2,2) followed by a Conv2D layer with 16 filters. A concatenate layer for concatenating with Conv2 along axis = 3
\item
Conv8: Same as Conv6 with 16 filters.
\item
Up9: UpSampling2D with size = (2,2) followed by a Conv2D layer with 8 filters. A concatenate layer for concatenating with Conv1 along axis = 3
\item
Conv9: Same as Conv6 with 8 filters
\item
Conv10: A Conv2D layers with 9 filters, stride = (1,1), activation = softmax
\end{enumerate}

We trained the network using the stochastic gradient descent method (Nesterov momentum=0.9, learning rate=0.001), and the model with the least validation loss was considered as the best model. The network was constructed using the deep learning library Keras (version 2.3) with Tensorflow (version 2.0) backend and then it was trained on an NVIDIA GTX 1080 founder's edition GPU with CUDA v8.0 and cuDNN v5.1 acceleration.

\section{Loss function}
\paragraph{}
The loss function ($L_{R}$) for quantifying image reconstruction loss was computed from the mean of the Jaccard index calculated for each tissue as:

\begin{align}
J_{i} & = \sum_{i=1}^{N} \frac{|P_{i} \cap M_{i}|}{|P_{i} \cup M_{i}|} \nonumber \\
L_{R} & = 1-\frac{J}{N}
 \label{eq:loss function}
\end{align}

where $J_{i}$ is the Jaccard Index for the tissue $i$, $N$ is the total number of classes in the manually segmented image, $P_{i}$ is the set of pixels belonging to class $i$ as predicted by the network, and $M_{i}$ is the set of pixels representing the class $i$ in the manual segmentation.

We used Categorical crossentropy loss ($L_{c}$) for the classification network. It is a loss function that is used in multiclass classification tasks, where an image can only belong to one out of many possible classes. The loss function can be calculated as:

\begin{align}
L_{C} & = \sum_{i=1}^{N}  y_{i} ~  log(\hat{y}_{i})
\label{eq:crossentropy loss function}
\end{align}

where $\hat{y}_{i}$ is the i-th scalar value in the model output, $y_{i}$ is the corresponding target value, and $N$ is the output size, i.e., the number of scalar values in the model output.

The total loss was the sum of both loss functions, i.e., $L = L_{R} +L_{C} $

\section{Data Augmentation}
\paragraph{}
We performed data augmentation to deal with the sparsity of the training data. For the segmentation network, data augmentation consisted of rotation ($\pm 4$ degrees), horizontal/ vertical translations ($\pm 10$ pixels), horizontal flipping, nonlinear intensity shift and multiplicative speckle noise (see the work of Devalla et al.for more details \cite{devalla2018drunet}).

For the autoencoder network, we only used rotation ($\pm 4$ degrees), horizontal/ vertical translations ($\pm 10$ pixels) for data augmentation.

\begin{figure}
     \centering
     \begin{subfigure}[b]{0.45\textwidth}
         \centering
         \includegraphics[height = 6cm, width = 7cm]{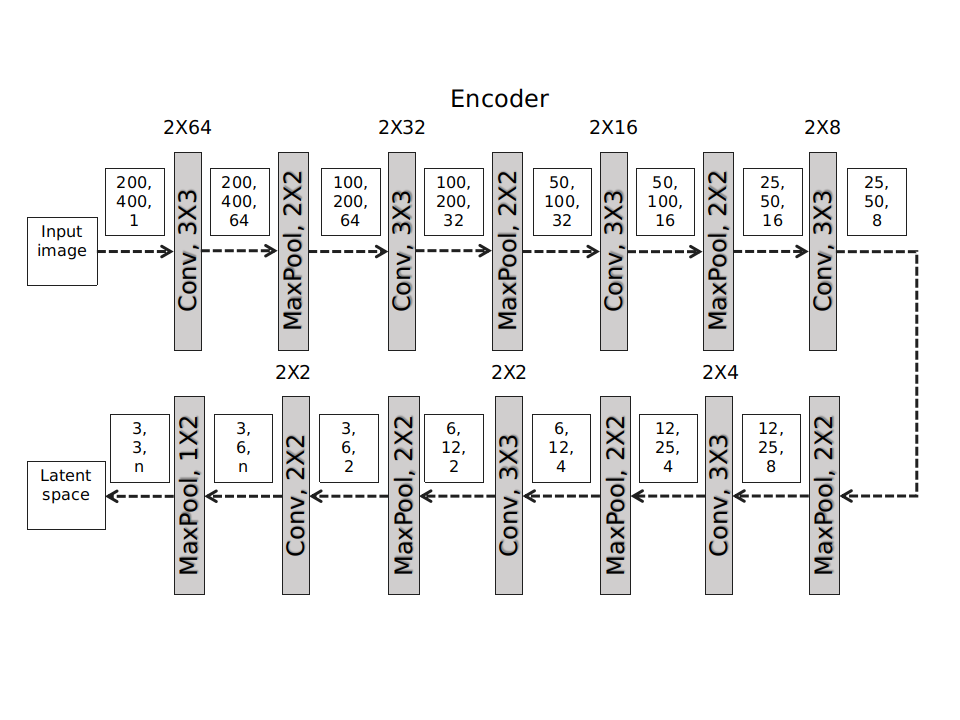}
         \caption{}
     \end{subfigure}
     \begin{subfigure}[b]{0.45\textwidth}
         \centering
         \includegraphics[height = 6cm, width = 7 cm]{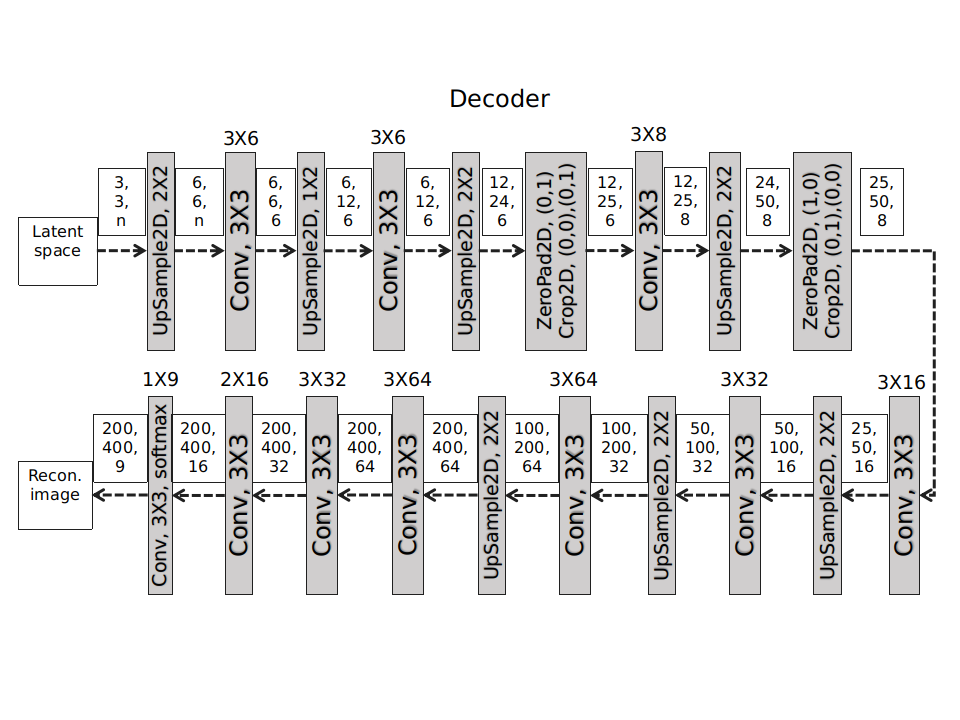}
         \caption{}
     \end{subfigure} \\
     \begin{subfigure}[b]{0.45\textwidth}
         \centering
         \includegraphics[height = 6cm, width = 7 cm]{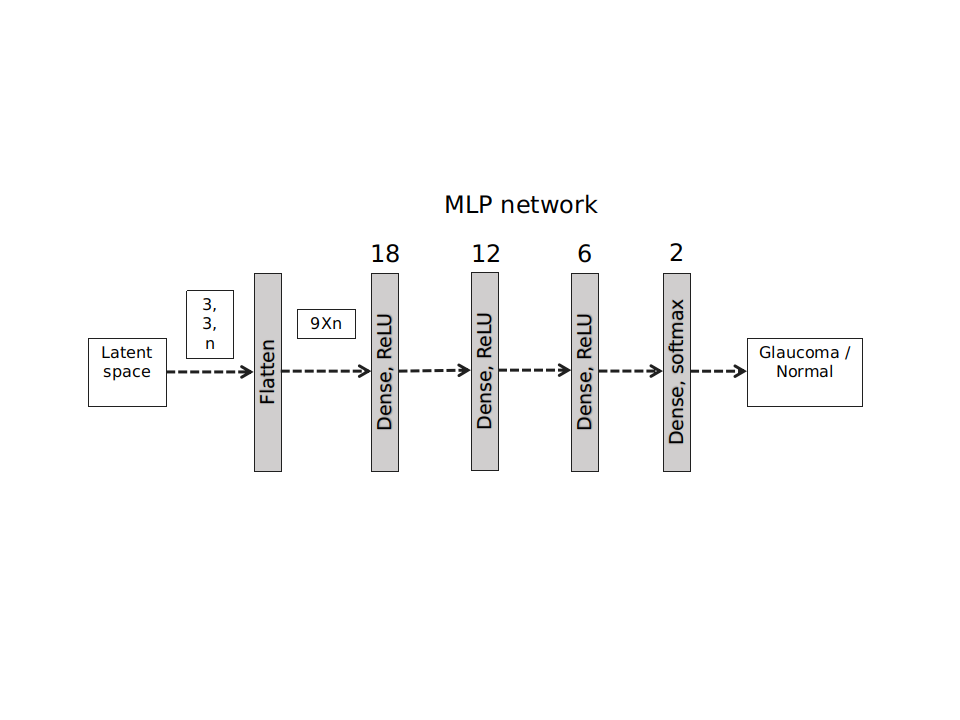}
         \caption{}
     \end{subfigure}
     \begin{subfigure}[b]{0.45\textwidth}
         \centering
         \includegraphics[height = 6cm, width = 7 cm]{Fig2.png}
         \caption{}
     \end{subfigure}
        \caption{ Autoencoder architecture }
\end{figure}

\section{Autoencoder architecture}
\paragraph{}
Both encoder and decoder used CNN layers. The encoder reduced the size of the input image ($200 \times 400$ pixels) to a lower dimension ($d$). $d=3 \times 3 \times n$, where $n$ is a user-defined variable to define the size of the latent space. We increased the value of $n$ linearly from one to eight, and thus, the latent space size was increased from 9 to 72.

\subsection*{Encoder architecture}

\begin{enumerate}
\item 
Conv1: Two 2D convolution (Conv2D) layers with 64 filters (stride = (3,3), padding = same,  kernel initializer = `he normal'), followed by a 2d max-pooling layer (MaxPooling2D) with pool size=(2, 2).
\item
Conv2: Same as Conv1 with 32 filter.
\item
Conv3: Same as Conv1 with 16 filter.
\item
Conv4: Same as Conv1 with 8 filter.
\item
Conv5: Same as Conv1 with 4 filter.
\item
Conv6: Same as Conv1 with 2 filter.
\item
Conv7: Two 2D convolution (Conv2D) layers with 2 filters (stride = (2,2), padding = same,  kernel initializer = `he normal'), followed by a 2d max-pooling layer (MaxPooling2D) with pool size=(1, 2).
\end{enumerate}

\subsection*{Decoder architecture}

\begin{enumerate}
\item 
Up1: UpSampling2D with size = (2,2) followed by three Conv2D layer with 6 filters (stride = (3,3), padding = same,  kernel initializer = `he normal').
\item 
Up2: UpSampling2D with size = (1,2) followed by three Conv2D layer with 6 filters (stride = (3,3), padding = same,  kernel initializer = `he normal'). 
\item 
Up3: UpSampling2D with size = (2,2) followed by ZeroPadding2D (0,1), Crop2D((0,0),(0,1)), followed by three Conv2D layer with 8 filters (stride = (3,3), padding = same,  kernel initializer = `he normal').
\item 
Up4: UpSampling2D with size = (2,2) followed by ZeroPadding2D (1,0), Crop2D((0,1),(0,0)), followed by three Conv2D layer with 16 filters (stride = (3,3), padding = same,  kernel initializer = `he normal').
\item
Up5: UpSampling2D with size = (2,2) followed by three Conv2D layer with 32 filters (stride = (3,3), padding = same,  kernel initializer = `he normal').
\item
Up6: Same as Up5 with three Conv2D layer with 64 filters
\item
Up7: Same as Up5 with three Conv2D layer with 64 filters
\item
Up8: three Conv2D layer with 32 filters, two Conv2D layer with 16 filters, one Conv2D layer with 9 filters, followed by a Softmax activation.
\end{enumerate}

\subsection*{MLP branch architecture}
\paragraph{}
A Flatten layer followed by four Dense layers with 18, 12, 6, and 2 neurons, followed by a Softmax activation.

\section{Matrices for performance comparison of autoencoders}
\paragraph{}
The receiver operating characteristic curve is a metric that illustrates the diagnostic ability of a binary classifier system. A value of one for AUC represents a perfect classification with $100\%$ sensitivity (no false negatives) and $100\%$ specificity (no false positives), whereas a value of 0.5 indicates random guesses.

\section{ONH structural information}
\paragraph{}

\begin{figure}[!ht]
	\centering
	\includegraphics[height = 6cm, width = 10cm]{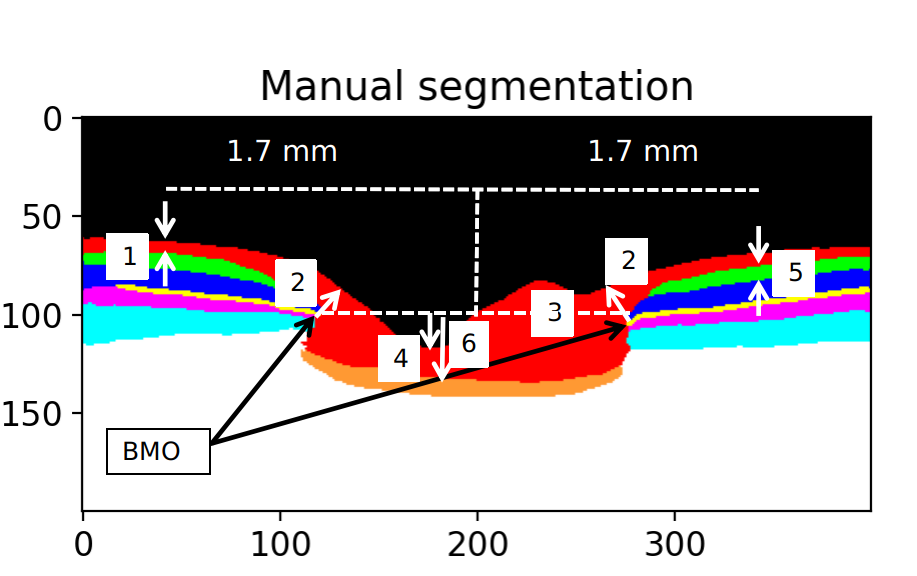}
	\caption{ONH tissue features. 1: RNFL thickness, 2: Minimum rim width (MRW), 3: Disc size, 4: Prelamina depth, 5: GCL+IPL thickness, 6: LC depth }
	\label{fig:OCT_terminology}
\end{figure}

\section{Variation in PC2 and PC3}
\paragraph{}

\begin{figure*}[htp]
	\centering
	\includegraphics[height = 7cm, width = 0.6\textwidth]{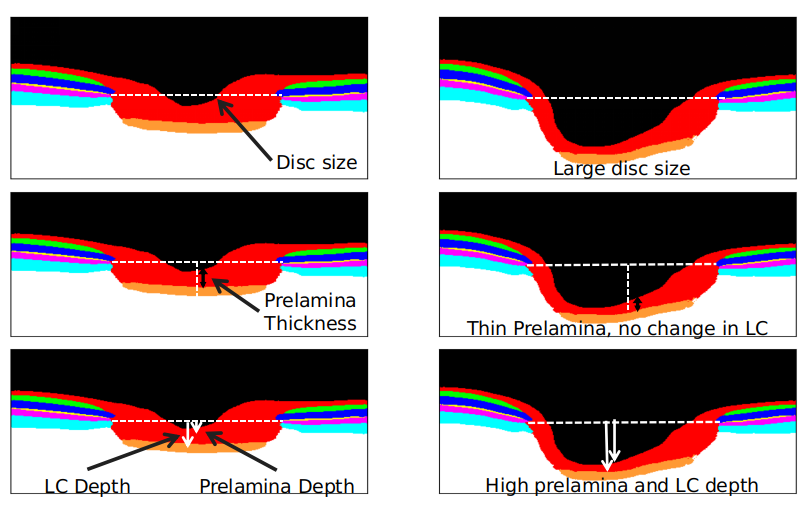}
	\caption{Structural changes in the ONH with variations in PC2.}
	\label{fig:PC2_Variation}
\end{figure*}

\begin{figure}[htp]
	\centering
	\includegraphics[height = 7cm, width = 0.6\textwidth]{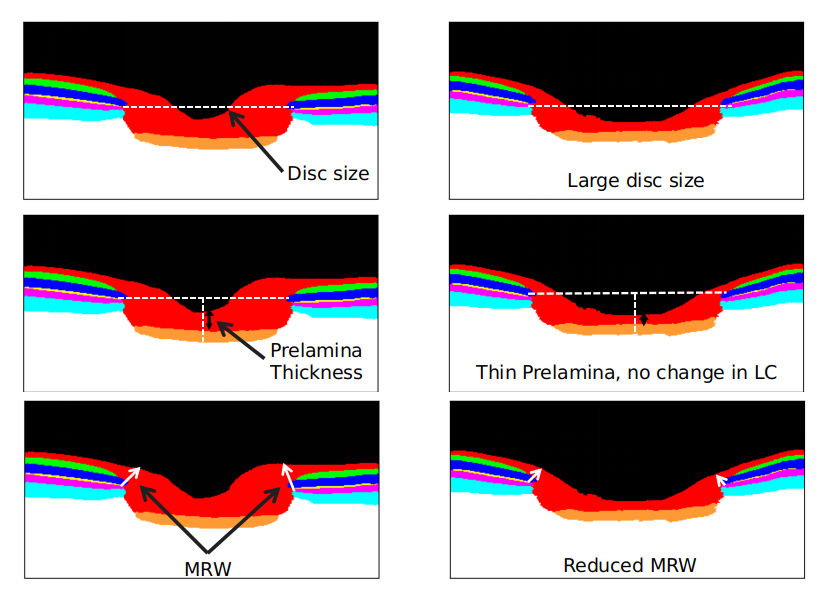}
	\caption{Structural changes in the ONH with variations in PC3.}
	\label{fig:PC3_Variation}
\end{figure}

\section{Variations in other PCs}
\paragraph{}

\begin{figure}
     \centering
     \begin{subfigure}[b]{0.9\textwidth}
         \centering
         \includegraphics[height = 2cm, width = 11cm]{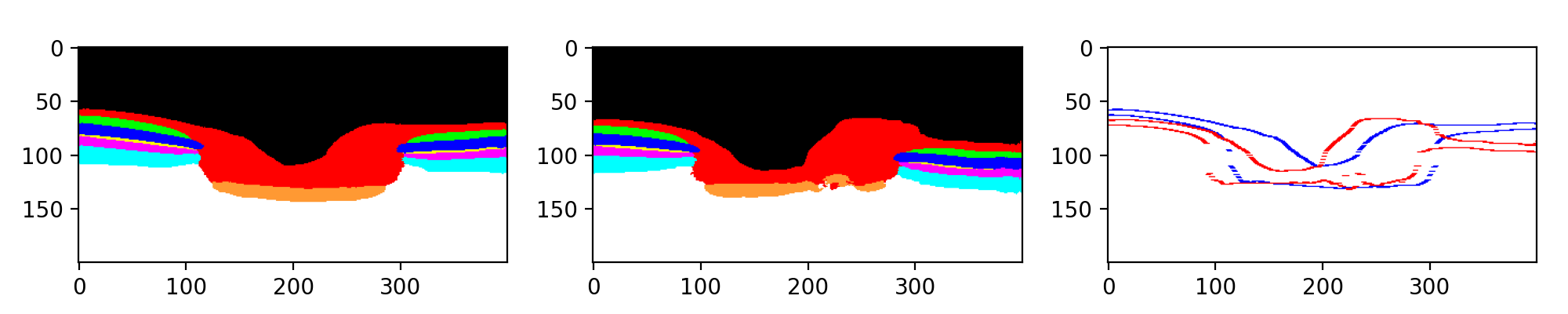}
         \caption{}
         \label{fig:OrgModEdgeComp_2}
     \end{subfigure}\\
     \begin{subfigure}[b]{0.9\textwidth}
         \centering
         \includegraphics[height = 2cm, width = 11 cm]{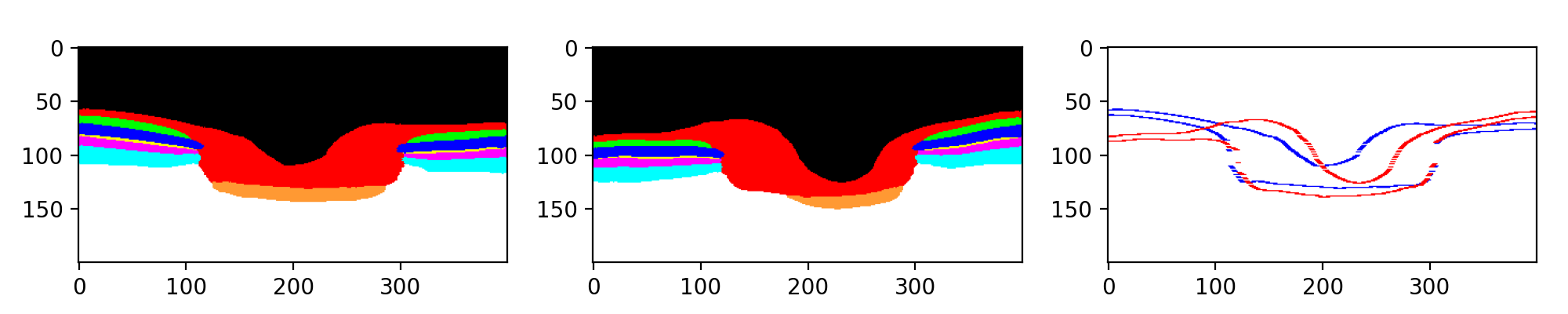}
         \caption{}
         \label{fig:OrgModEdgeComp_3}
     \end{subfigure} \\
     \begin{subfigure}[b]{0.9\textwidth}
         \centering
         \includegraphics[height = 2cm, width = 11 cm]{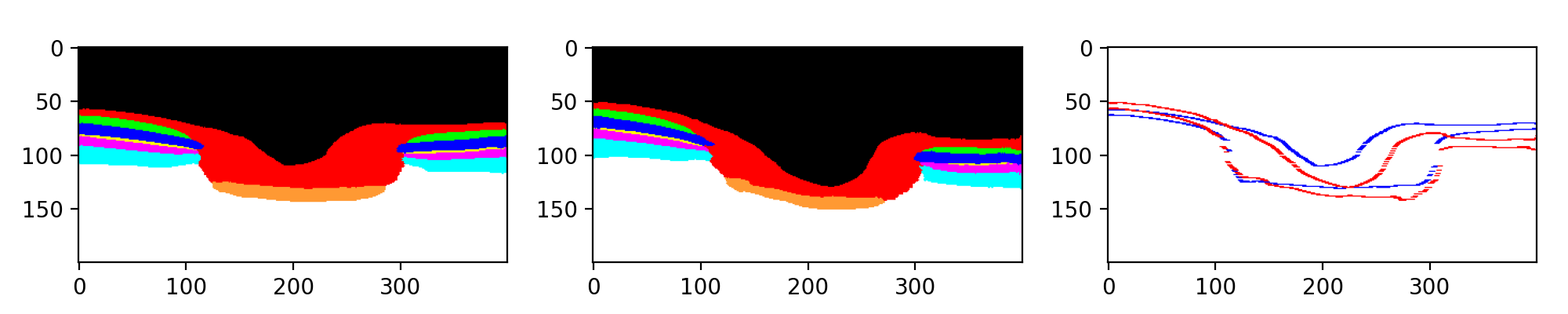}
         \caption{}
         \label{fig:OrgModEdgeComp_4}
     \end{subfigure} \\
     \begin{subfigure}[b]{0.9\textwidth}
         \centering
         \includegraphics[height = 2cm, width = 11 cm]{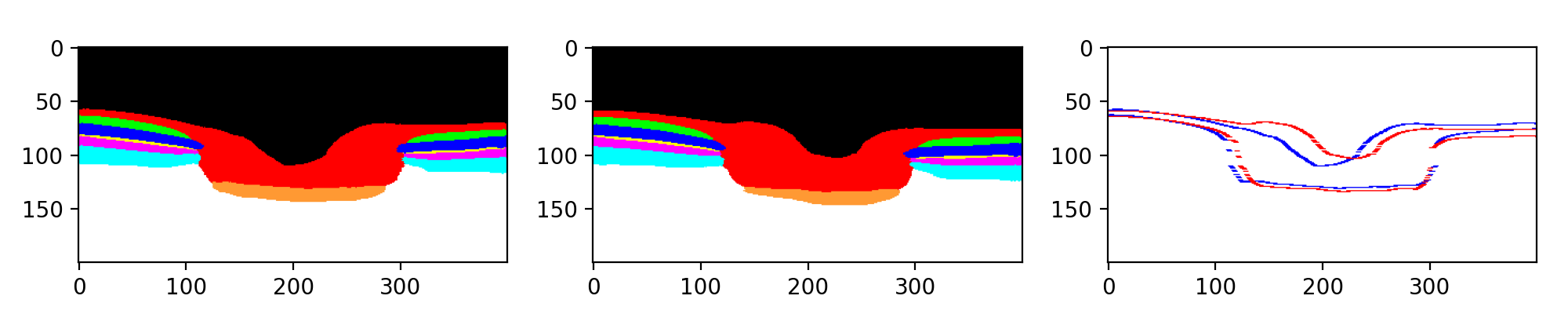}
         \caption{}
         \label{fig:OrgModEdgeComp_5}
     \end{subfigure} \\
     \begin{subfigure}[b]{0.9\textwidth}
         \centering
         \includegraphics[height = 2cm, width = 11 cm]{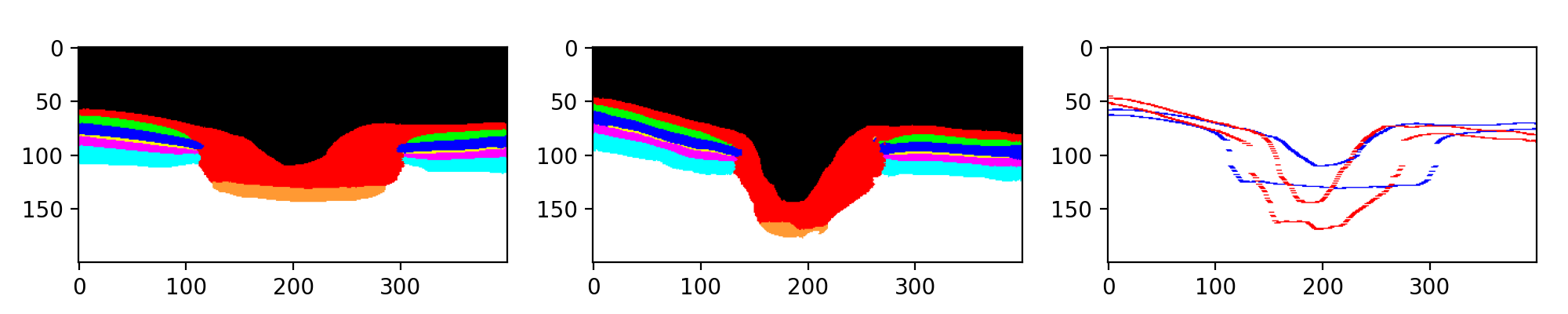}
         \caption{}
         \label{fig:OrgModEdgeComp_7}
     \end{subfigure} \\
     \begin{subfigure}[b]{0.9\textwidth}
         \centering
         \includegraphics[height = 2cm, width = 11 cm]{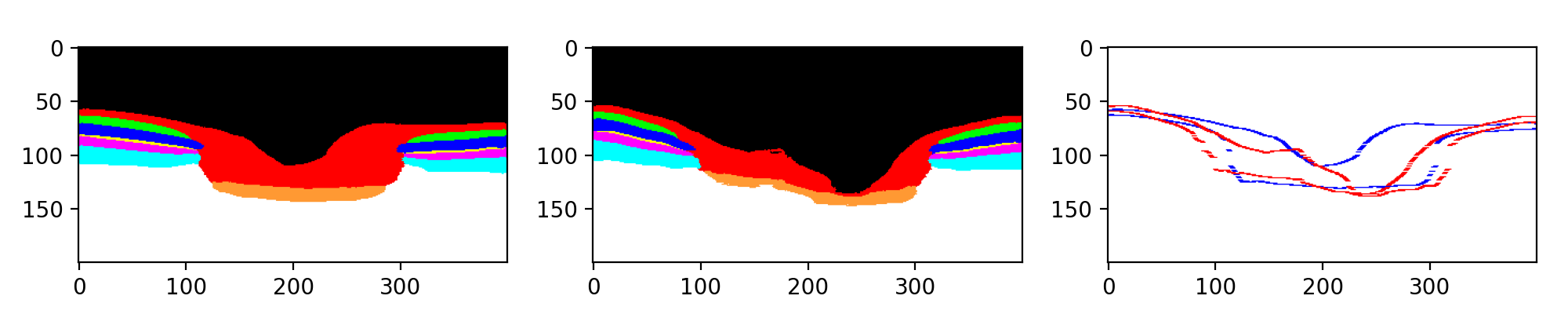}
         \caption{}
         \label{fig:OrgModEdgeComp_8}
     \end{subfigure} \\
     \begin{subfigure}[b]{0.9\textwidth}
         \centering
         \includegraphics[height = 2cm, width = 11 cm]{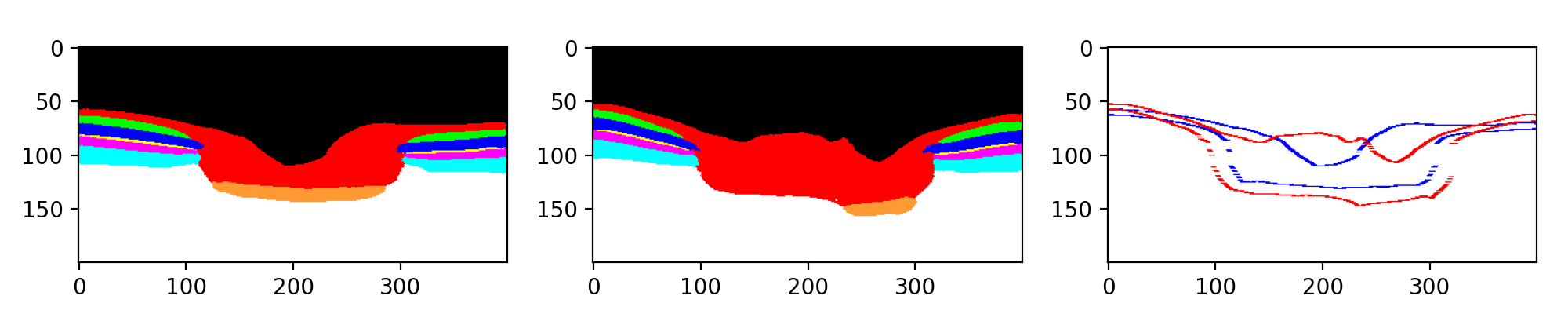}
         \caption{}
         \label{fig:OrgModEdgeComp_9}
     \end{subfigure} 
     \caption{Changes in the ONH tissue with variations in PC4-10   }
     \label{fig:change with other pc 4 to 10}
\end{figure}

Fig.\ref{fig:change with other pc 4 to 10} shows the changes in the ONH tissue structure with variation in PC4 to PC10 components.

\begin{enumerate}
\item
Fig.\ref{fig:OrgModEdgeComp_2}: Variation in PC4. This PC represented the thickening of the right-hand side (RHS) MRW with a twisting of the RHS ONH tissue. Also, thinning of the left-hand side (LHS) MRW was noticed.
\item
Fig.\ref{fig:OrgModEdgeComp_3}: Variation in PC5. This PC represented the thickening of the LHS MRW with a twisting of the LHS ONH tissue and thinning of RHS MRW.
\item
Fig.\ref{fig:OrgModEdgeComp_4}: Variation in PC6. Thinning of the prelamina tissue. 
\item
Fig.\ref{fig:OrgModEdgeComp_5}: Variation in PC7. Thickening of the prelamina tissue. 
\item
Fig.\ref{fig:OrgModEdgeComp_7}: Variation in PC8. Reduction in disc size. 
\item
Fig.\ref{fig:OrgModEdgeComp_8}: Variation in PC9. Prelamina thinning with distortion. Also, an increase in disc size.
\item
Fig.\ref{fig:OrgModEdgeComp_9}: Variation in PC10. Thickening of prelamina with vanishing LC.
\end{enumerate}

We were unable to notice any significant change in the ONH region with variations in the other PCs.

\section{Results with unsupervised autoencoder}
\paragraph{}
The architecture of the unsupervised autoencoder is presented in Fig.\ref{fig:Autoencoder_unsupervised}. The encoder and the decoder architectures were the same as that in Fig.\ref{fig:Autoencoder}. Fig.\ref{fig:AutoencoderDiceCoefficient_unsupervized} shows the Dice coefficient on test images. The mean value was $0.88 \pm 0.03$. Fig.\ref{fig:PCA_clustering_54_unsupervized} and Fig.\ref{fig:UMAP_clustering_54_unsupervized} show the latent features after PCA and UMAP transformation. The unsupervised autoencoder was not able to differentiate non-glaucoma and glaucomatous eyes correctly.

\begin{figure}
     \centering
     \begin{subfigure}[b]{0.45\textwidth}
         \centering
         \includegraphics[height = 6cm, width = 8cm]{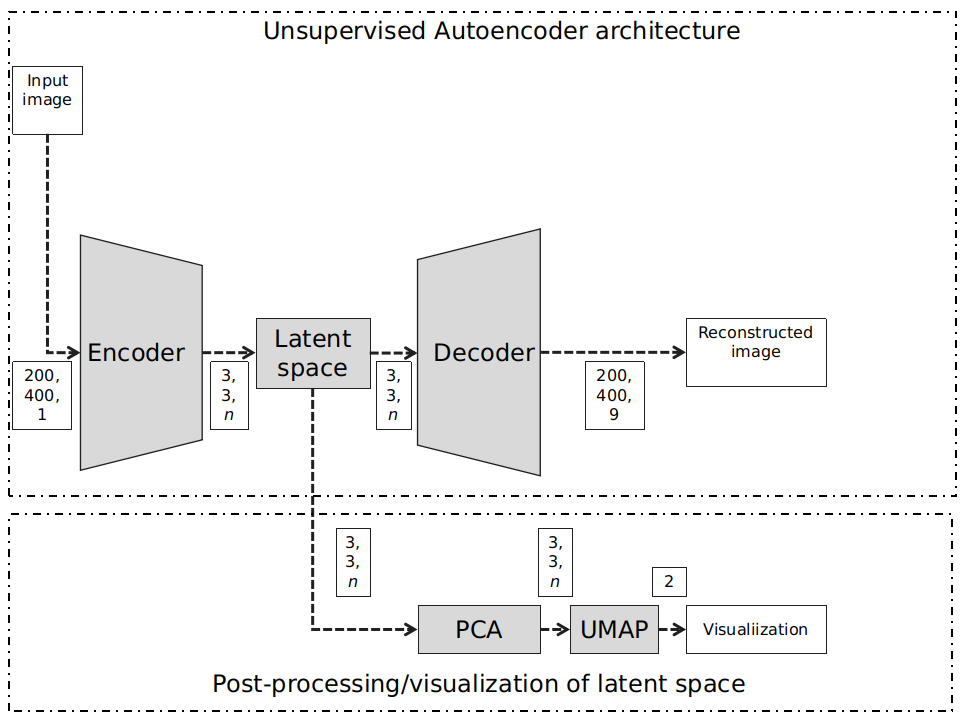}
         \caption{}
         \label{fig:Autoencoder_unsupervised}
     \end{subfigure}
     \begin{subfigure}[b]{0.45\textwidth}
         \centering
         \includegraphics[height = 6cm, width = 3 cm]{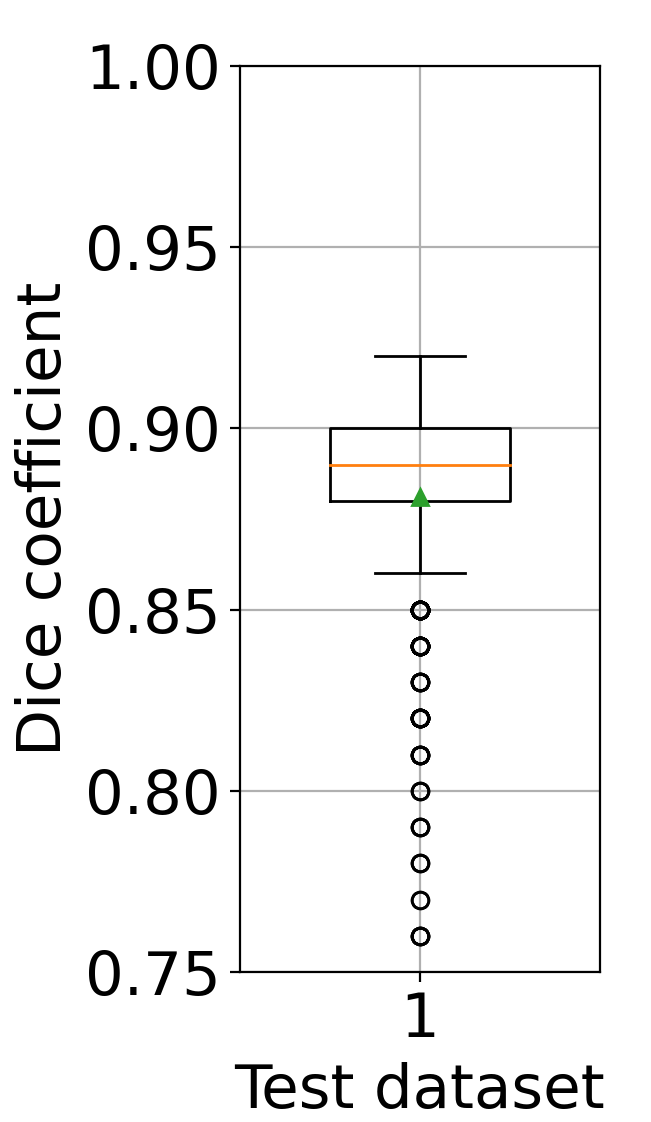}
         \caption{}
         \label{fig:AutoencoderDiceCoefficient_unsupervized}
     \end{subfigure} \\
     \begin{subfigure}[b]{0.45\textwidth}
         \centering
         \includegraphics[height = 6cm, width = 8 cm]{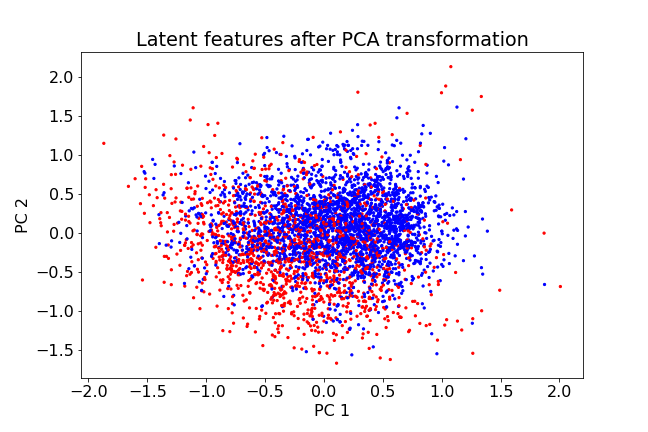}
         \caption{}
         \label{fig:PCA_clustering_54_unsupervized}
     \end{subfigure}
     \begin{subfigure}[b]{0.45\textwidth}
         \centering
         \includegraphics[height = 6cm, width = 8 cm]{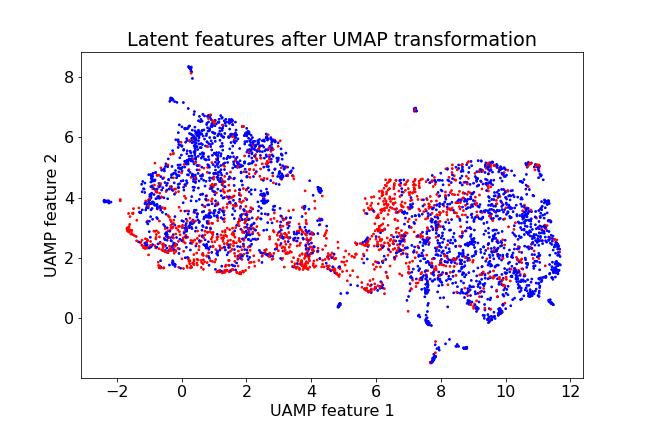}
         \caption{}
         \label{fig:UMAP_clustering_54_unsupervized}
     \end{subfigure}
     \label{fig:unsupervised learning}
        \caption{ Results of unsupervised learning. a) The network architecture. b) The distribution of the Dice coefficients with 1111 test images as a box plot. c) Visualization of PC1 and PC2 features as a scatter plot after PCA transformation. d) Visualization of UMAP features as a scatter plot after UMAP transformation. Blue points: non-glaucoma, and red points: glaucoma } 
\end{figure}

\end{small}

\end{document}